\documentclass[superscriptaddress,showpacs,amsmath,amssymb,twocolumn,nofootinbib]{revtex4-2}

\usepackage{amsmath, amsthm, amssymb, bm, braket}
\usepackage[dvips]{graphicx}
\usepackage{color}
\newcommand{\tomo}[1]{\textcolor[rgb]{0.0,0.0,0.0}{#1}}

\begin{document}
\title{Spin correlation in two-proton emission from $^6$Be}

\author{Tomohiro Oishi}
\email[E-mail: ]{tomohiro.oishi@ribf.riken.jp}
\affiliation{RIKEN Nishina Center for Accelerator-Based Science, Wako 351-0198, Japan}

\renewcommand{\figurename}{FIG.}
\renewcommand{\tablename}{TABLE}

\newcommand{\bi}[1]{\ensuremath{\boldsymbol{#1}}}
\newcommand{\unit}[1]{\ensuremath{\mathrm{#1}}}
\newcommand{\oprt}[1]{\ensuremath{\hat{\mathcal{#1}}}}
\newcommand{\abs}[1]{\ensuremath{\left| #1 \right|}}
\newcommand{\slashed}[1] {\not\!{#1}} 
\newcommand{\crc}[1] {c^{\dagger}_{#1}}
\newcommand{\anc}[1] {c_{#1}}
\newcommand{\crb}[1] {\alpha^{\dagger}_{#1}}
\newcommand{\anb}[1] {\alpha_{#1}}

\def \beq{\begin{equation}}
\def \eeq{\end{equation}}
\def \beqa{\begin{eqnarray}}
\def \eeqa{\end{eqnarray}}
\def \bir{\bi{r}}
\def \ubir{\bar{\bi{r}}}
\def \bip{\bi{p}}
\def \ubip{\bar{\bi{r}}}
\def \adel{\tilde{l}}
\def \twop{$2p$}

\begin{abstract}
{\noindent
This paper presents a theoretical evaluation of spin correlation in the two-proton ($2p$) radioactive emission.
The three-body model of $^{6}$Be with the proton-proton interaction, which is adjusted to reproduce the experimental energy release, is utilized.
Time-dependent calculation is performed to compute the coupled-spin state of the emitted two protons.
The spin-correlation function $S$ as the Clauser-Horne-Shimony-Holt (CHSH) indicator is evaluated as $|S| \cong 2.65$.
Namely, the $2p$-spin correlation beyond the limit of local-hidden-variable (LHV) theory is suggested.
This correlation is sensitive to the proton-proton interaction.
The short-lived (broad-width) $2p$~state has the weaker spin correlation.
In parallel, the core-proton interactions do not harm this correlation during the time-dependent decaying process.
The CHSH measurement can be a novel probe into the effective nuclear interaction inside finite systems.
Two-proton emitters can provide a testing field for identical-particle entanglement.
}
\end{abstract}

\maketitle

\section{introduction}
Clauser-Horne-Shimony-Holt (CHSH) inequality is one essential property of the quantum-entangled state \cite{1969CHSH}.
This is one variant of Bell inequality introduced by John Clauser {\it et. al.} 
for the proof of Bell theorem, which claims that certain consequence of quantum mechanics cannot be reproduced by the local-hidden-variable (LHV) theory \cite{64Bell,04Bell}.
By using the CHSH indicator $S$, the limit of LHV theory is symbolically given as $\abs{S} \leq 2$.
Throughout the history of Bell-CHSH examinations \cite{1978Clauser,1982Aspect,1998Weihs,2000JianWei,2001Rowe,2010Scheidl,2015Giustina,2015Shalm,2017Vasilyev,2023Storz}, the violation of LHV-theory limit has been confirmed.
In these examinations, the entangled states of photons, electrons, and atoms are populated and measured to satisfy $\abs{S} > 2$.
I emphasize that, for this purpose, many efforts have been devoted to close the loopholes, including those of detection \cite{2001Rowe,2015Giustina,2015Shalm}, locality \cite{1982Aspect,1998Weihs,2010Scheidl}, and memory \cite{2002Barrett}.

In the nuclear physics, the quantum entanglement can play a role in various
scenes \cite{1976Rachti, 2003Sakai, 2015Kanada, 2023Bulgac_PRC,2023Pazy,2023Gu,2023Heng,2023Obiol, 2024Kou, 2024Krichner,2021Robin,2021Kruppa,2023Miller,2023Miller_PRCLett,2023Tichai,2023Sun_PRC,2023Johnson,2024Bai}.
In Ref. \cite{1976Rachti} by Lamehi-Rachti and Mittig, the low-energy proton-proton scattering was measured for testing the Bell inequality.
In this pioneer work, some extra assumptions were required to certify the violation of LHV-theory limit.
In Ref. \cite{2003Sakai}, Sakai {\it et. al.} performed a novel measurement to demonstrate that a strong \tomo{spin correlation} is realized between two protons.
This experiment measures the spin-singlet two protons made in the reaction of $^2$H+$p \longrightarrow ^2$He+$n$.
The spin-correlation function as the CHSH indicator is deduced as $S_{\rm expt} = 2.83 \pm 0.24_{\rm stat} \pm 0.07_{\rm sys}$, which is in agreement with the non-local quantum mechanics and beyond the LHV-theory limit.
Several theoretical works have been devoted to compute the entanglement in atomic nuclei \cite{2015Kanada, 2023Bulgac_PRC,2023Pazy,2023Gu,2023Heng,2023Obiol, 2024Kou, 2024Krichner,2021Robin,2021Kruppa,2023Miller,2023Miller_PRCLett,2023Tichai,2023Sun_PRC,2023Johnson,2024Bai}.
In Ref. \cite{2023Bulgac_PRC}, due to the nuclear short-range correlations, the occupation probabilities of nuclear orbits change to increase the entanglement entropy.
In Ref. \cite{2024Bai}, the evaluation of nuclear spin entanglement with the quantum-state tomography is suggested to be feasible.
In Refs. \cite{2024Bai,2023Johnson,2023Miller_PRCLett}, a finite proton-neutron entanglement is suggested.

Another possible example to observe the nuclear \tomo{spin correlation} is the two-proton (\twop) radioactive emission \cite{08Bertulani,2009Gri_rev,2012Pfu_rev,08Blank_01,08Blank_02,2023Pfutzner_rev,2019Qi_rev}.
In this radioactive decay, the parent nuclei spontaneously decay by emitting two protons.
Especially in so-called ``prompt'' \twop~emission \cite{2009Gri_rev, 60Gold, 61Gold},
the two protons are expected to have the diproton-like clustering and/or the dominant spin-singlet configuration.
This is attributable to the effective \twop~interaction inside finite systems, being in a contrast to the vacuum \twop~interaction supporting no bound states.

\tomo{For identical particles, a careful interpretation of correlation and entanglement is
necessary \cite{2020Benatti,2021Frederick,2001Paskauskas,2002Eckert,2002Ghirardi,2004Ghirardi,2003Shi,2004Barnum,2004Zanardi,2011Sasaki,2013Balacha,2015Reusch,2011Tichy,2016Franco,2017Benatti,2020Morris,2020Farzam,2023Frerot,2023Ptaszynski,2024Serrano}.
For distinguishable particles, if their wave function is factorized to one direct product, no correlation exists, and their entanglement is trivially zero.
The same discussion, however, cannot apply to a pair of protons, namely, identical fermions.
Because of the anti-symmetrization, its wave function should not be factorized.
Although many studies accumulated \cite{2020Benatti,2021Frederick},
no consensus on the well-defined entanglement in such a condition has been obtained.
One category of protocols is to focus only on the correlations between observable quantities \cite{2004Barnum,2004Zanardi,2011Sasaki,2015Reusch,2013Balacha}.
Another idea is to introduce a new definition of entanglement for identical particles \cite{2001Paskauskas,2002Eckert,2002Ghirardi,2004Ghirardi,2003Shi,2011Sasaki}.
Recent studies suggest that the identical-particle entanglement can be utilized as resource for quantum operations \cite{2020Benatti,2016Franco,2020Morris}.
For future debates of the identical-particle entanglement,
in this work, I focus on whether a \twop-spin correlation beyond the LHV-theory limit can exist or not.
}

I evaluate the \tomo{spin correlation} in the \twop~emission from the $0^+$ state of $^{6}$Be \cite{88Ajzen,91Ajzen,1989Boch,09Gri_80,09Gri_677,12Ego}.
This $^{6}$Be is the lightest \twop~emitter, and well approximated by a simple three-body model with time dependence \cite{2014Oishi,2017Oishi,2021Wang_Naza}.
For the two protons spontaneously emitted, the measurement of spin correlation for the CHSH examination is assumed \cite{1969CHSH,2003Sakai}.
Because of the three-body problem, the \twop~state is under the effect of the third particle, namely, the daughter $^{4}$He nucleus.
Whether this effect destroys or not the \tomo{correlation} is investigated.
Its sensitivity to the proton-proton interaction is also discussed.

\section{Formalism and Model}
\begin{figure}[t] \begin{center}
\includegraphics[width = 0.95\hsize]{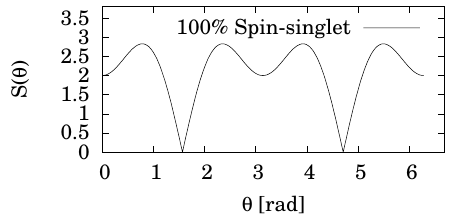}
\caption{CHSH indicator of the pure spin-singlet state.}
\label{fig:PEBIRE}
\end{center} \end{figure}

I consider the coupling of two protons, i.e., identical spin-$1/2$ fermions, and simulate the CHSH examination.
First, the four options of measurement by the two observers, so-called ``Alice'' and ``Bob'' conventionally, are introduced:
\beqa
&&
\hat{A}_{1,\theta} (1) \otimes \hat{B}_{1,\theta}(2),~~
\hat{A}_{2,\theta} (1) \otimes \hat{B}_{1,\theta}(2),
\nonumber \\
&&
\hat{A}_{1,\theta} (1) \otimes \hat{B}_{2,\theta}(2),~~
\hat{A}_{2,\theta} (1) \otimes \hat{B}_{2,\theta}(2). \label{eq:options}
\eeqa
Namely, Alice observes the first fermion with one chosen from the two options, $\hat{A}_{1,\theta}$ and $\hat{A}_{2,\theta}$.
Bob does the second fermion with one chosen from $\hat{B}_{1,\theta}$ and $\hat{B}_{2,\theta}$.
Those operators including the parameter angle $\theta$ are given as
\beq
\begin{array}{l}
  \hat{A}_{1,\theta} (1) = \hat{\sigma}_z(1), \\
  \hat{A}_{2,\theta} (1) = \hat{\sigma}_z(1) \cos 2\theta +\hat{\sigma}_x(1) \sin 2\theta,
\end{array}
\eeq
for Alice, whereas
\beq
\begin{array}{l}
  \hat{B}_{1,\theta} (2) = \hat{\sigma}_z(2) \cos \theta +\hat{\sigma}_x(2) \sin \theta, \\
  \hat{B}_{2,\theta} (2) = \hat{\sigma}_z(2) \cos \theta -\hat{\sigma}_x(2) \sin \theta,
\end{array}
\eeq
for Bob.
For an arbitrary two-fermion state $\ket{\Psi(1,2)}$, their expectation values are obtained:
\beq
\Braket{A_i B_j} = \Braket{ \Psi(1,2)  \mid \hat{A}_{i,\theta} (1) \otimes \hat{B}_{j,\theta}(2) \mid  \Psi(1,2)}.
\eeq
Then the CHSH indicator $S$ is determined as
\beq
S = \max  \left\{  \abs{S_{-+++}},\abs{S_{+-++}},\abs{S_{++-+}},\abs{S_{+++-}}  \right\},
\eeq
where
\beqa
S_{-+++} &=& -\Braket{A_1 B_1}+\Braket{A_2 B_1}+\Braket{A_1 B_2}+\Braket{A_2 B_2}, \nonumber \\
S_{+-++} &=& \Braket{A_1 B_1}-\Braket{A_2 B_1}+\Braket{A_1 B_2}+\Braket{A_2 B_2}, \nonumber \\
S_{++-+} &=& \Braket{A_1 B_1}+\Braket{A_2 B_1}-\Braket{A_1 B_2}+\Braket{A_2 B_2}, \nonumber \\
S_{+++-} &=& \Braket{A_1 B_1}+\Braket{A_2 B_1}+\Braket{A_1 B_2}-\Braket{A_2 B_2}. \label{eq:CHSHs}
\eeqa
In FIG. \ref{fig:PEBIRE}, $S$ is calculated for the spin-singlet state $\ket{\uparrow \downarrow -\downarrow \uparrow}/\sqrt{2}$.
At $\theta=\pi/4,~3\pi/4,~5\pi/4$, and $7\pi/4$, it has the maximum value, $S=2\sqrt{2}$ (Tsirelson's bound) \cite{1980Tsirelson}.
This situation resembles Sakai's experiment \cite{2003Sakai}.
In the following sections, $\theta=\pi/4$.
As well known, when the initial state is one of the Bell states $\left\{ \ket{e_n}  \right\}$, one finds $S=2\sqrt{2}$.
These states read \cite{2000Nielsen}
\beqa
&& \ket{e_1}= \frac{1}{\sqrt{2}} \ket{\uparrow \uparrow +\downarrow \downarrow},~~~~~
\ket{e_2}= \frac{i}{\sqrt{2}} \ket{\uparrow \uparrow -\downarrow \downarrow}, \nonumber \\
&& \ket{e_3}= \frac{i}{\sqrt{2}} \ket{\uparrow \downarrow +\downarrow \uparrow},~~~~~
\ket{e_4}= \frac{1}{\sqrt{2}} \ket{\uparrow \downarrow -\downarrow \uparrow}.
\eeqa
They can be used as basis for an arbitrary total-spin state of two spin-$1/2$ fermions.

\tomo{In this work, instead of $\left\{ \ket{e_n }  \right\}$,
the total-spin eigenstates $\left\{ \ket{d_k } \right\}$ are utilized for calculations \cite{60Edm}.}
Namely, for the total spin $\hat{S}_{12}$, their eigenvalues are given as $\hat{S}^2_{12}\ket{d_k} =S(S+1) \ket{d_k}$
and $\hat{S}_{12,z}\ket{d_k} = V\ket{d_k}$, where
\beqa
\ket{d_1} &=& \ket{S=1,V=+1} =\ket{\uparrow \uparrow}, \nonumber \\
\ket{d_2} &=& \ket{S=1,V=-1}=\ket{\downarrow \downarrow}, \nonumber \\
\ket{d_3} &=& \ket{S=1,V=0} =\frac{1}{\sqrt{2}} \ket{\uparrow \downarrow +\downarrow \uparrow}, \nonumber \\
\ket{d_4} &=& \ket{S=0,V=0} =\frac{1}{\sqrt{2}} \ket{\uparrow \downarrow -\downarrow \uparrow}.
\eeqa
\tomo{Notice that $\left\{ \ket{d_k } \right\}$ and $\left\{ \ket{e_n } \right\}$ are unitary transformable.}
\tomo{In addition, the actual \twop-wave function has the coordinate degrees of freedom.
Thus,
\beqa
\Psi (\bir_1 \sigma_1 ,\bir_2 \sigma_2) &=& \Braket{ \bir_1 \sigma_1 ,\bir_2 \sigma_2 \mid \Psi(1,2)}  \nonumber  \\
&=& \sum_{k=1}^{4} F_k(\bir_1,\bir_2) \Braket{\sigma_1 ,\sigma_2 \mid d_k}. \label{eq:FFFFFF}
\eeqa
This wave function is originally expanded on the anti-symmetrized \twop~basis \cite{2014Oishi,2017Oishi}.
Thus, $\Psi (\bir_2 \sigma_2 ,\bir_1 \sigma_1) = (-)\Psi (\bir_1 \sigma_1 ,\bir_2 \sigma_2)$.
For computing the expectation values in Eq. (\ref{eq:CHSHs}), these coordinate parts are integrated as
\beqa
\Braket{A_i B_j} &=& \sum_{l,k} a_{lk} \Braket{d_l \mid A_i B_j \mid d_k},   \\
a_{lk} &=& \iint d\bir_1 d\bir_2 F^*_l (\bir_1,\bir_2) F_k (\bir_1,\bir_2). \label{eq:alk}
\eeqa
In actual computations, $F_k (\bir_1,\bir_2)$ as well as $a_{lk}$ are numerically solved, whereas $\Braket{d_l \mid A_i B_j \mid d_k}$ can be analytically evaluated.
}

I employ the three-body model, which has been developed and utilized in Refs. \cite{88Suzuki_COSM,1991BE,1997EBH,2005HS,2014Oishi,2017Oishi}.
The system contains the $^{4}$He nucleus as the rigid core with mass $m_C$ and two valence protons.
The two valence protons feel the spherical mean field $V(\bir)$ generated by the $^{4}$He core.
Thus, the three-body Hamiltonian reads
\beqa
&& \hat{H}_{3B} = \hat{h}(\bir_1)+\hat{h}(\bir_2) +v_{pp}(\bir_1,\bir_2) +\frac{\bip_1 \cdot \bip_2}{m_C}, \label{eq:H3B} \\
&& \hat{h}(r_i) = -\frac{\hbar^2}{2\mu} \frac{d^2}{dr_i^2} +\frac{\hbar^2}{2\mu} \frac{l(l+1)}{r_i^2} + V(r_i),
\eeqa
where $\mu = m_p m_C /(m_p + m_C)$ and $m_p=938.272$ MeV$/c^2$ for protons.
\tomo{By diagonalizing $\hat{H}_{3B}$, the anti-symmetrized \twop-wave function
$\Psi (\bir_1 \sigma_1 ,\bir_2 \sigma_2)$ is obtained \cite{2014Oishi,2017Oishi}.}
Here $\hat{h}(\bir_k)$ is of the $k$th proton-core subsystem.
For its interaction $V(r_i)$, the same Woods-Saxon and Coulomb potentials in the previous work \cite{2017Oishi} are employed.
The proton-proton interaction reads
\beq
v_{pp}(\bir_1,\bir_2) = v_{nucl}(d) + \frac{e^2}{4\pi \epsilon_0} \frac{1}{d} +v_{add}(\bir_1,\bir_2),
\eeq
where $d=\abs{\bir_2 -\bir_1}$.
The nuclear-force term is described by the spin-dependent Gaussian potential \cite{77Thom}:
\beqa
v_{nucl}(d) &=& \left[ V_R e^{-a_R d^2}  +V_S e^{-a_S d^2} \right] \hat{P}_{S=0} \nonumber  \\
&& +\left[ V_R e^{-a_R d^2}  +V_T e^{-a_T d^2} \right] \hat{P}_{S=1}.
\eeqa
The operator $\hat{P}_{S=0}$ ($\hat{P}_{S=1}$) indicates the projection into the
spin-singlet (spin-triplet) channel of the proton-proton subsystem.
Parameters are given as
$V_R=200$ MeV,
$V_S=-91.85$ MeV,
$V_T=-178$ MeV,
$a_R=1,487$ fm$^{-2}$,
$a_S=0.465$ fm$^{-2}$, and
$a_T=0.639$ fm$^{-2}$.
These parameters correctly reproduce the experimental vacuum-scattering length of two protons \cite{77Thom}.
In addition, the surface-dependent term $v_{add}$ is employed.
That is,
\beq
v_{add} (\bir_1,\bir_2) = w_0 e^{-(R-R_0)^2 /B_0^2} \delta (\bir_1-\bir_2), \label{eq:vppadd}
\eeq
where $R=\abs{(\bir_1 +\bir_2) /2}$,
$R_0 = 1.68$ fm,
$B_0 = 0.6 R_0$, and
$w_0 = -470$ MeV$\cdot$fm$^3$.
This additional term is necessary to reproduce the experimental energy:
the \twop~energy and width are obtained as $E_{2p}=1.356$ MeV and $\Gamma_{2p}=0.055$ MeV, respectively,
whereas the experimental data read $E_{2p}=1.372(5)$ MeV and $\Gamma_{2p} = 0.092(6)$ MeV \cite{88Ajzen,91Ajzen}.
Notice that this additional potential vanishes when one of the three particles is infinitely separated.
Thus, the vacuum properties of two-body subsystems can be conserved in the time-development calculations.
Note that the emitted two protons are unbound in vacuum.

\begin{figure}[t] \begin{center}
\includegraphics[width = 0.95\hsize]{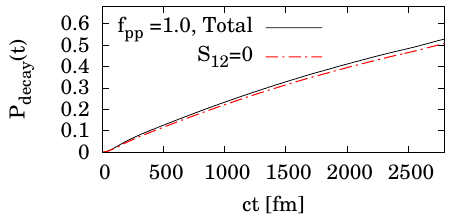}
\caption{
Time-dependent decaying probability of \twop~emission from $^{6}$Be.
Its spin-singlet ($S_{12}=0$) component is also plotted.
The factor $f_{pp}=1.0$ indicates that the proton-proton interaction is adjusted to reproduce the experimental \twop-energy.
}
\label{fig:2024_0705}
\end{center} \end{figure}

\section{Results and Discussions}
The \twop-emitting process is simulated with the time-dependent method \cite{2014Oishi,2017Oishi}:
\beq
\ket{\Psi(t)} = \exp \left[ -it \frac{\hat{H}_{3B}}{\hbar} \right] \ket{\Psi(0)}, \label{eq:TDPP}
\eeq
where the initial state is solved as the confined \twop~state inside the Coulomb barrier \cite{2014Oishi}.
The decaying state $\ket{\Psi_d(t)}$, which describes the emitted component outside the barrier,
is determined as
\beq
\ket{\Psi_d(t)} = \ket{\Psi(t)}  -\beta(t) \ket{\Psi(0)},
\eeq
where $\beta(t)$ is the survival coefficient, 
$\beta(t)=\Braket{\Psi(0) | \Psi(t)}$.

FIG. \ref{fig:2024_0705} displays the time-dependent decaying probability \cite{2014Oishi,2017Oishi}:
\beq
P_{decay}(t) = \Braket{\Psi_d(t) \mid \Psi_d(t)} = 1-\abs{\beta(t)}^2 .
\eeq
One clearly finds that $P_{decay}(t)$ increases in time development.
The spin-singlet state, $\ket{d_4} = \ket{e_4} =\ket{\uparrow \downarrow -\downarrow \uparrow}/\sqrt{2}$, is always dominant.
I confirmed that the survival probability,
\beq
P_{surv}(t) \equiv \Braket{\Psi(0) \mid \Psi(t)}^2 = \abs{\beta(t)}^2 = 1-P_{decay}(t),
\eeq
is well approximated by the exponential damping as $P_{surv}(t) \cong e^{-t/\tau}$, where the width reads $\Gamma_{2p}=\hbar / \tau \cong 0.055$ MeV by numerical fitting.
Also, by observing the decaying density distribution, $\rho_{d}(t,\bir_1,\bir_2) = \abs{ \Psi_d(t,\bir_1,\bir_2)}^2$,
I confirmed that this process can be interpreted as the diproton-correlating emission \cite{2014Oishi,2017Oishi}.

\begin{figure}[t] \begin{center}
\includegraphics[width = 0.95\hsize]{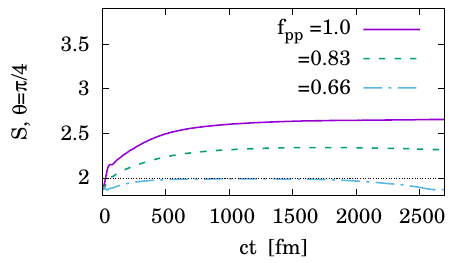}
\caption{
CHSH indicator of the time-dependent decaying state $\ket{\Psi_d(t)}$ of the \twop~emission of $^{6}$Be~$\longrightarrow ^{4}$He$+p+p$.
Here $\theta = \pi/4$.
The factor $f_{pp}$ indicates the strength of the proton-proton interaction.
}\label{fig:2024_0624}
\end{center} \end{figure}

\begin{table}[b] \begin{center}
\caption{Two-proton energy, width, and the CHSH indicator of $^{6}$Be for several $f_{pp}$ values.
The corresponding lifetime, $\tau = \hbar /\Gamma_{2p}$, is also displayed.
The CHSH indicator $S$ is evaluated at $ct=1600$ fm.
Experimental data read $E_{2p}=1.372(5)$ MeV and $\Gamma_{2p} = 0.092(6)$ MeV \cite{88Ajzen,91Ajzen}.
}
\label{table:KIM}
\catcode`? = \active \def?{\phantom{0}} 
  \begingroup \renewcommand{\arraystretch}{1.2}
  \begin{tabular*}{\hsize} { @{\extracolsep{\fill}} cccc}
\hline \hline
 $f_{pp}$   &$E_{2p}$~[MeV]  &$\Gamma_{2p}$~[MeV]~~~($\tau$~[s]) &$S$ \\  \hline
 $1.00$   &$1.356$         &$0.055$~~($1.2 \times 10^{-20}$)   &$2.65$   \\
 $0.83$   &$2.056$         &$0.177$~~($3.7 \times 10^{-21}$)   &$2.34$   \\
 $0.66$   &$2.616$         &$0.428$~~($1.5 \times 10^{-21}$)   &$1.98$   \\
\hline \hline
  \end{tabular*}
  \endgroup
  \catcode`? = 12 
\end{center} \end{table}

In Fig. \ref{fig:2024_0624}, the CHSH indicator $S$ is evaluated for the decaying state by using the expansion in Eq. (\ref{eq:FFFFFF}):
\beq
\Psi_d(t,\bir_1 \sigma_1 ,\bir_2 \sigma_2) = \sum_{k=1}^{4} F_k (t,\bir_1,\bir_2) \Braket{\sigma_1 ,\sigma_2 \mid d_k}.
\eeq
Clearly $S>2$, i.e. beyond the LHV-theory limit, during the time evolution.
For $ct \geq 1500$ fm, I obtain $S \cong 2.65$ with the default setting of $v_{pp}$ ($f_{pp}=1.0$) for reproducing the experimental $E_{2p}$.
This spin correlation is attributable to the dominant spin-singlet component in FIG \ref{fig:2024_0705}.
Notice also that the present time evolution in Eq. (\ref{eq:TDPP}) contains the effect of core-proton interactions $V(r_i)$.
These interactions do not harm the spin correlation.

For deeper understanding, the sensitivity of CHSH indicator $S$ to the proton-proton interaction is studied.
For this purpose, a tuning factor is employed.
That is,
\beq
v_{pp}(\bir_1,\bir_2) \longrightarrow f_{pp} v_{pp}(\bir_1,\bir_2).
\eeq
With $f_{pp}=1$, it reproduces the experimental energy, $E_{2p}=1.356$ MeV \cite{88Ajzen,91Ajzen}.
\tomo{Note that $v_{pp}$ is not tunable in real nuclear experiments, and thus, this is purely a numerical test.
In FIG. \ref{fig:2024_0624} and TABLE \ref{table:KIM}, my results are summarized.
FIG. \ref{fig:2024_0624} shows that $S$ becomes reduced when $v_{pp}$ is weakened.
Especially, it goes below the LHV-theory limit, $S<2$, with $f_{pp} \leq 0.66$.
From TABLE \ref{table:KIM}, $E_{2p}$ and $\Gamma_{2p}$ noticeably depend on $f_{pp}$.
The short-lived (broad-width) \twop~state has the smaller spin correlation.
Consequently, even in the identical-\twop~case without factorizability, if their interaction was weak, the CHSH examination could fail to reject the LHV theory.}

\begin{figure}[t] \begin{center}
\includegraphics[width = 0.95\hsize]{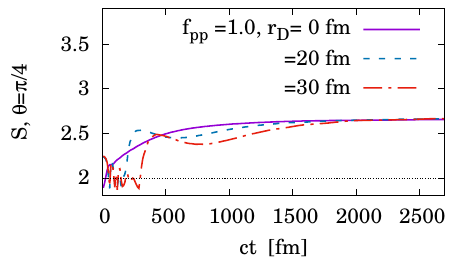}
\caption{
Same to the $f_{pp}=1.0$ result in FIG. \ref{fig:2024_0624}, but by changing the detecting radius $r_D$: see Eq. (\ref{eq:alk_rd}).
}\label{fig:2025_0106}
\end{center} \end{figure}

\tomo{In experiments, the emitted two protons are detected after distances from the daughter nucleus.
By using the detecting radius $r_D$, Eq. (\ref{eq:alk}) is modified as
\beq
a_{lk}(r_D) = \iint_{\abs{\bir_i} \geq r_D} d\bir_1 d\bir_2 F^*_l (\bir_1,\bir_2) F_k (\bir_1,\bir_2). \label{eq:alk_rd}
\eeq
In FIG. \ref{fig:2025_0106}, the CHSH indicator by changing $r_D$ is displayed.
Except the short-time fluctuations, the result converges as $S =2.65$ independently of $r_D$.
In parallel, in actual experiments, $r_D$ can be of the order of $10^{12}$ fm or more.
Such a setting goes beyond the present computing capability.
This problem is left for future improvement.
}

\section{Summary}
In this paper, the \tomo{\twop-spin correlation} is evaluated for the \twop~emission from $^{6}$Be.
As a product of proton-proton interaction, $S \cong 2.65$ is obtained in the time-dependent decaying state: \tomo{a correlation} beyond the LHV-theory limit is suggested.
This \tomo{correlation} is not harmed by the core-proton interactions.
The CHSH measurement can be a probe into the effective nuclear interaction inside finite systems.

This work is limited to the lightest $^{6}$Be nucleus.
Possibility of spin correlation in other \twop~emitters is an open question.
For evaluating the entanglement, von-Neumann entropy has been considered as an essential
quantity \cite{2015Kanada,2023Bulgac_PRC,2023Pazy,2023Gu,2023Heng,2023Obiol,2024Kou,2024Krichner},
whereas this work focuses only on the spin correlation.
\tomo{In Ref. \cite{2009Horod}, a modern definition based on the quantum information theory is proposed.
There, entangled states are defined as those, which cannot be generated by a sequence including only local operations and classical communications (LOCC).}
These topics will be addressed in forthcoming studies.

\tomo{For the identical-\twop~state, whether its spin correlation can validate the proper entanglement or not is still under discussions \cite{2020Benatti,2021Frederick,2001Paskauskas,2002Eckert,2002Ghirardi,2004Ghirardi,2003Shi,2004Barnum,2004Zanardi,2011Sasaki,2013Balacha,2015Reusch,2011Tichy,2016Franco,2017Benatti,2020Morris,2020Farzam,2023Frerot,2023Ptaszynski,2024Serrano}.
Atomic nuclei can be a testing field for identical and distinguishable-particle (proton-neutron) entanglements \cite{2023Johnson,2023Miller_PRCLett,2024Bai} in equal scales.
}

In the experimental side, mass production of \twop~emitters, including the $^{6}$Be, for a sufficient statistics is still challenging.
For measuring the two protons by independent detectors after decay, an advanced design of experiment is necessary.
Closing all the loopholes of
detection \cite{2001Rowe,2015Giustina,2015Shalm},
locality \cite{1982Aspect,1998Weihs,2010Scheidl}, and
memory \cite{2002Barrett}
should require another lot of efforts.
On the other side, the experimental survey of \twop~emitters is rapidly in progress, e.g. that in RIKEN RIBF.
Confirmation of spin entanglement in \twop~radioactivity can be one landmark in the nuclear physics as well as quantum many-body science.

\begin{acknowledgments}
Numerical calculations are supported by the Multi-disciplinary Cooperative Research Program (MCRP) in FY2023 and FY2024 by Center for Computational Sciences, University of Tsukuba (project ID wo23i034), allocating computational resources of supercomputer Wisteria/BDEC-01 (Odyssey) in Information Technology Center, University of Tokyo.
We appreciate the cooperative project of supercomputer Yukawa-21 in Yukawa Institute for Theoretical Physics, Kyoto University.
\end{acknowledgments}

%


\begin{thebibliography}{79}%
\makeatletter
\providecommand \@ifxundefined [1]{%
 \@ifx{#1\undefined}
}%
\providecommand \@ifnum [1]{%
 \ifnum #1\expandafter \@firstoftwo
 \else \expandafter \@secondoftwo
 \fi
}%
\providecommand \@ifx [1]{%
 \ifx #1\expandafter \@firstoftwo
 \else \expandafter \@secondoftwo
 \fi
}%
\providecommand \natexlab [1]{#1}%
\providecommand \enquote  [1]{``#1''}%
\providecommand \bibnamefont  [1]{#1}%
\providecommand \bibfnamefont [1]{#1}%
\providecommand \citenamefont [1]{#1}%
\providecommand \href@noop [0]{\@secondoftwo}%
\providecommand \href [0]{\begingroup \@sanitize@url \@href}%
\providecommand \@href[1]{\@@startlink{#1}\@@href}%
\providecommand \@@href[1]{\endgroup#1\@@endlink}%
\providecommand \@sanitize@url [0]{\catcode `\\12\catcode `\$12\catcode
  `\&12\catcode `\#12\catcode `\^12\catcode `\_12\catcode `\%12\relax}%
\providecommand \@@startlink[1]{}%
\providecommand \@@endlink[0]{}%
\providecommand \url  [0]{\begingroup\@sanitize@url \@url }%
\providecommand \@url [1]{\endgroup\@href {#1}{\urlprefix }}%
\providecommand \urlprefix  [0]{URL }%
\providecommand \Eprint [0]{\href }%
\providecommand \doibase [0]{https://doi.org/}%
\providecommand \selectlanguage [0]{\@gobble}%
\providecommand \bibinfo  [0]{\@secondoftwo}%
\providecommand \bibfield  [0]{\@secondoftwo}%
\providecommand \translation [1]{[#1]}%
\providecommand \BibitemOpen [0]{}%
\providecommand \bibitemStop [0]{}%
\providecommand \bibitemNoStop [0]{.\EOS\space}%
\providecommand \EOS [0]{\spacefactor3000\relax}%
\providecommand \BibitemShut  [1]{\csname bibitem#1\endcsname}%
\let\auto@bib@innerbib\@empty
\bibitem [{\citenamefont {Clauser}\ \emph {et~al.}(1969)\citenamefont
  {Clauser}, \citenamefont {Horne}, \citenamefont {Shimony},\ and\
  \citenamefont {Holt}}]{1969CHSH}%
  \BibitemOpen
  \bibfield  {author} {\bibinfo {author} {\bibfnamefont {J.~F.}\ \bibnamefont
  {Clauser}}, \bibinfo {author} {\bibfnamefont {M.~A.}\ \bibnamefont {Horne}},
  \bibinfo {author} {\bibfnamefont {A.}~\bibnamefont {Shimony}},\ and\ \bibinfo
  {author} {\bibfnamefont {R.~A.}\ \bibnamefont {Holt}},\ }\href
  {https://doi.org/10.1103/PhysRevLett.23.880} {\bibfield  {journal} {\bibinfo
  {journal} {Phys. Rev. Lett.}\ }\textbf {\bibinfo {volume} {23}},\ \bibinfo
  {pages} {880} (\bibinfo {year} {1969})}\BibitemShut {NoStop}%
\bibitem [{\citenamefont {Bell}(1964)}]{64Bell}%
  \BibitemOpen
  \bibfield  {author} {\bibinfo {author} {\bibfnamefont {J.}~\bibnamefont
  {Bell}},\ }\href@noop {} {\bibfield  {journal} {\bibinfo  {journal}
  {Physics}\ }\textbf {\bibinfo {volume} {1}},\ \bibinfo {pages} {195}
  (\bibinfo {year} {1964})}\BibitemShut {NoStop}%
\bibitem [{\citenamefont {Bell}(2004)}]{04Bell}%
  \BibitemOpen
  \bibfield  {author} {\bibinfo {author} {\bibfnamefont {J.~S.}\ \bibnamefont
  {Bell}},\ }\href@noop {} {\emph {\bibinfo {title} {Speakable and Unspeakable
  in Quantum Mechanics}}},\ \bibinfo {edition} {2nd}\ ed.,\ Collected Papers on
  Quantum Philosophy\ (\bibinfo  {publisher} {Cambridge University Press},\
  \bibinfo {address} {Cambridge, UK},\ \bibinfo {year} {2004})\BibitemShut
  {NoStop}%
\bibitem [{\citenamefont {Clauser}\ and\ \citenamefont
  {Shimony}(1978)}]{1978Clauser}%
  \BibitemOpen
  \bibfield  {author} {\bibinfo {author} {\bibfnamefont {J.~F.}\ \bibnamefont
  {Clauser}}\ and\ \bibinfo {author} {\bibfnamefont {A.}~\bibnamefont
  {Shimony}},\ }\href {https://doi.org/10.1088/0034-4885/41/12/002} {\bibfield
  {journal} {\bibinfo  {journal} {Reports on Progress in Physics}\ }\textbf
  {\bibinfo {volume} {41}},\ \bibinfo {pages} {1881} (\bibinfo {year}
  {1978})}\BibitemShut {NoStop}%
\bibitem [{\citenamefont {Aspect}\ \emph {et~al.}(1982)\citenamefont {Aspect},
  \citenamefont {Grangier},\ and\ \citenamefont {Roger}}]{1982Aspect}%
  \BibitemOpen
  \bibfield  {author} {\bibinfo {author} {\bibfnamefont {A.}~\bibnamefont
  {Aspect}}, \bibinfo {author} {\bibfnamefont {P.}~\bibnamefont {Grangier}},\
  and\ \bibinfo {author} {\bibfnamefont {G.}~\bibnamefont {Roger}},\ }\href
  {https://doi.org/10.1103/PhysRevLett.49.91} {\bibfield  {journal} {\bibinfo
  {journal} {Phys. Rev. Lett.}\ }\textbf {\bibinfo {volume} {49}},\ \bibinfo
  {pages} {91} (\bibinfo {year} {1982})}\BibitemShut {NoStop}%
\bibitem [{\citenamefont {Weihs}\ \emph {et~al.}(1998)\citenamefont {Weihs},
  \citenamefont {Jennewein}, \citenamefont {Simon}, \citenamefont
  {Weinfurter},\ and\ \citenamefont {Zeilinger}}]{1998Weihs}%
  \BibitemOpen
  \bibfield  {author} {\bibinfo {author} {\bibfnamefont {G.}~\bibnamefont
  {Weihs}}, \bibinfo {author} {\bibfnamefont {T.}~\bibnamefont {Jennewein}},
  \bibinfo {author} {\bibfnamefont {C.}~\bibnamefont {Simon}}, \bibinfo
  {author} {\bibfnamefont {H.}~\bibnamefont {Weinfurter}},\ and\ \bibinfo
  {author} {\bibfnamefont {A.}~\bibnamefont {Zeilinger}},\ }\href
  {https://doi.org/10.1103/PhysRevLett.81.5039} {\bibfield  {journal} {\bibinfo
   {journal} {Phys. Rev. Lett.}\ }\textbf {\bibinfo {volume} {81}},\ \bibinfo
  {pages} {5039} (\bibinfo {year} {1998})}\BibitemShut {NoStop}%
\bibitem [{\citenamefont {Pan}\ \emph {et~al.}(2000)\citenamefont {Pan},
  \citenamefont {Bouwmeester}, \citenamefont {Daniell}, \citenamefont
  {Weinfurter},\ and\ \citenamefont {Zeilinger}}]{2000JianWei}%
  \BibitemOpen
  \bibfield  {author} {\bibinfo {author} {\bibfnamefont {J.-W.}\ \bibnamefont
  {Pan}}, \bibinfo {author} {\bibfnamefont {D.}~\bibnamefont {Bouwmeester}},
  \bibinfo {author} {\bibfnamefont {M.}~\bibnamefont {Daniell}}, \bibinfo
  {author} {\bibfnamefont {H.}~\bibnamefont {Weinfurter}},\ and\ \bibinfo
  {author} {\bibfnamefont {A.}~\bibnamefont {Zeilinger}},\ }\href
  {https://doi.org/10.1038/35000514} {\bibfield  {journal} {\bibinfo  {journal}
  {Nature}\ }\textbf {\bibinfo {volume} {403}},\ \bibinfo {pages} {515}
  (\bibinfo {year} {2000})}\BibitemShut {NoStop}%
\bibitem [{\citenamefont {Rowe}\ \emph {et~al.}(2001)\citenamefont {Rowe},
  \citenamefont {Kielpinski}, \citenamefont {Meyer}, \citenamefont {Sackett},
  \citenamefont {Itano}, \citenamefont {Monroe},\ and\ \citenamefont
  {Wineland}}]{2001Rowe}%
  \BibitemOpen
  \bibfield  {author} {\bibinfo {author} {\bibfnamefont {M.~A.}\ \bibnamefont
  {Rowe}}, \bibinfo {author} {\bibfnamefont {D.}~\bibnamefont {Kielpinski}},
  \bibinfo {author} {\bibfnamefont {V.}~\bibnamefont {Meyer}}, \bibinfo
  {author} {\bibfnamefont {C.~A.}\ \bibnamefont {Sackett}}, \bibinfo {author}
  {\bibfnamefont {W.~M.}\ \bibnamefont {Itano}}, \bibinfo {author}
  {\bibfnamefont {C.}~\bibnamefont {Monroe}},\ and\ \bibinfo {author}
  {\bibfnamefont {D.~J.}\ \bibnamefont {Wineland}},\ }\href
  {https://doi.org/10.1038/35057215} {\bibfield  {journal} {\bibinfo  {journal}
  {Nature}\ }\textbf {\bibinfo {volume} {409}},\ \bibinfo {pages} {791}
  (\bibinfo {year} {2001})}\BibitemShut {NoStop}%
\bibitem [{\citenamefont {Scheidl}\ \emph {et~al.}(2010)\citenamefont
  {Scheidl}, \citenamefont {Ursin}, \citenamefont {Kofler}, \citenamefont
  {Ramelow}, \citenamefont {Ma}, \citenamefont {Herbst}, \citenamefont
  {Ratschbacher}, \citenamefont {Fedrizzi}, \citenamefont {Langford},
  \citenamefont {Jennewein},\ and\ \citenamefont {Zeilinger}}]{2010Scheidl}%
  \BibitemOpen
  \bibfield  {author} {\bibinfo {author} {\bibfnamefont {T.}~\bibnamefont
  {Scheidl}}, \bibinfo {author} {\bibfnamefont {R.}~\bibnamefont {Ursin}},
  \bibinfo {author} {\bibfnamefont {J.}~\bibnamefont {Kofler}}, \bibinfo
  {author} {\bibfnamefont {S.}~\bibnamefont {Ramelow}}, \bibinfo {author}
  {\bibfnamefont {X.-S.}\ \bibnamefont {Ma}}, \bibinfo {author} {\bibfnamefont
  {T.}~\bibnamefont {Herbst}}, \bibinfo {author} {\bibfnamefont
  {L.}~\bibnamefont {Ratschbacher}}, \bibinfo {author} {\bibfnamefont
  {A.}~\bibnamefont {Fedrizzi}}, \bibinfo {author} {\bibfnamefont {N.~K.}\
  \bibnamefont {Langford}}, \bibinfo {author} {\bibfnamefont {T.}~\bibnamefont
  {Jennewein}},\ and\ \bibinfo {author} {\bibfnamefont {A.}~\bibnamefont
  {Zeilinger}},\ }\href {https://doi.org/10.1073/pnas.1002780107} {\bibfield
  {journal} {\bibinfo  {journal} {Proceedings of the National Academy of
  Sciences}\ }\textbf {\bibinfo {volume} {107}},\ \bibinfo {pages} {19708}
  (\bibinfo {year} {2010})},\ \Eprint
  {https://arxiv.org/abs/https://www.pnas.org/doi/pdf/10.1073/pnas.1002780107}
  {https://www.pnas.org/doi/pdf/10.1073/pnas.1002780107} \BibitemShut {NoStop}%
\bibitem [{\citenamefont {Giustina}\ \emph {et~al.}(2015)\citenamefont
  {Giustina}, \citenamefont {Versteegh}, \citenamefont {Wengerowsky},
  \citenamefont {Handsteiner}, \citenamefont {Hochrainer}, \citenamefont
  {Phelan}, \citenamefont {Steinlechner}, \citenamefont {Kofler}, \citenamefont
  {Larsson}, \citenamefont {Abell\'an}, \citenamefont {Amaya}, \citenamefont
  {Pruneri}, \citenamefont {Mitchell}, \citenamefont {Beyer}, \citenamefont
  {Gerrits}, \citenamefont {Lita}, \citenamefont {Shalm}, \citenamefont {Nam},
  \citenamefont {Scheidl}, \citenamefont {Ursin}, \citenamefont {Wittmann},\
  and\ \citenamefont {Zeilinger}}]{2015Giustina}%
  \BibitemOpen
  \bibfield  {author} {\bibinfo {author} {\bibfnamefont {M.}~\bibnamefont
  {Giustina}}, \bibinfo {author} {\bibfnamefont {M.~A.~M.}\ \bibnamefont
  {Versteegh}}, \bibinfo {author} {\bibfnamefont {S.}~\bibnamefont
  {Wengerowsky}}, \bibinfo {author} {\bibfnamefont {J.}~\bibnamefont
  {Handsteiner}}, \bibinfo {author} {\bibfnamefont {A.}~\bibnamefont
  {Hochrainer}}, \bibinfo {author} {\bibfnamefont {K.}~\bibnamefont {Phelan}},
  \bibinfo {author} {\bibfnamefont {F.}~\bibnamefont {Steinlechner}}, \bibinfo
  {author} {\bibfnamefont {J.}~\bibnamefont {Kofler}}, \bibinfo {author}
  {\bibfnamefont {J.-A.}\ \bibnamefont {Larsson}}, \bibinfo {author}
  {\bibfnamefont {C.}~\bibnamefont {Abell\'an}}, \bibinfo {author}
  {\bibfnamefont {W.}~\bibnamefont {Amaya}}, \bibinfo {author} {\bibfnamefont
  {V.}~\bibnamefont {Pruneri}}, \bibinfo {author} {\bibfnamefont {M.~W.}\
  \bibnamefont {Mitchell}}, \bibinfo {author} {\bibfnamefont {J.}~\bibnamefont
  {Beyer}}, \bibinfo {author} {\bibfnamefont {T.}~\bibnamefont {Gerrits}},
  \bibinfo {author} {\bibfnamefont {A.~E.}\ \bibnamefont {Lita}}, \bibinfo
  {author} {\bibfnamefont {L.~K.}\ \bibnamefont {Shalm}}, \bibinfo {author}
  {\bibfnamefont {S.~W.}\ \bibnamefont {Nam}}, \bibinfo {author} {\bibfnamefont
  {T.}~\bibnamefont {Scheidl}}, \bibinfo {author} {\bibfnamefont
  {R.}~\bibnamefont {Ursin}}, \bibinfo {author} {\bibfnamefont
  {B.}~\bibnamefont {Wittmann}},\ and\ \bibinfo {author} {\bibfnamefont
  {A.}~\bibnamefont {Zeilinger}},\ }\href
  {https://doi.org/10.1103/PhysRevLett.115.250401} {\bibfield  {journal}
  {\bibinfo  {journal} {Phys. Rev. Lett.}\ }\textbf {\bibinfo {volume} {115}},\
  \bibinfo {pages} {250401} (\bibinfo {year} {2015})}\BibitemShut {NoStop}%
\bibitem [{\citenamefont {Shalm}\ \emph {et~al.}(2015)\citenamefont {Shalm},
  \citenamefont {Meyer-Scott}, \citenamefont {Christensen}, \citenamefont
  {Bierhorst}, \citenamefont {Wayne}, \citenamefont {Stevens}, \citenamefont
  {Gerrits}, \citenamefont {Glancy}, \citenamefont {Hamel}, \citenamefont
  {Allman}, \citenamefont {Coakley}, \citenamefont {Dyer}, \citenamefont
  {Hodge}, \citenamefont {Lita}, \citenamefont {Verma}, \citenamefont
  {Lambrocco}, \citenamefont {Tortorici}, \citenamefont {Migdall},
  \citenamefont {Zhang}, \citenamefont {Kumor}, \citenamefont {Farr},
  \citenamefont {Marsili}, \citenamefont {Shaw}, \citenamefont {Stern},
  \citenamefont {Abell\'an}, \citenamefont {Amaya}, \citenamefont {Pruneri},
  \citenamefont {Jennewein}, \citenamefont {Mitchell}, \citenamefont {Kwiat},
  \citenamefont {Bienfang}, \citenamefont {Mirin}, \citenamefont {Knill},\ and\
  \citenamefont {Nam}}]{2015Shalm}%
  \BibitemOpen
  \bibfield  {author} {\bibinfo {author} {\bibfnamefont {L.~K.}\ \bibnamefont
  {Shalm}}, \bibinfo {author} {\bibfnamefont {E.}~\bibnamefont {Meyer-Scott}},
  \bibinfo {author} {\bibfnamefont {B.~G.}\ \bibnamefont {Christensen}},
  \bibinfo {author} {\bibfnamefont {P.}~\bibnamefont {Bierhorst}}, \bibinfo
  {author} {\bibfnamefont {M.~A.}\ \bibnamefont {Wayne}}, \bibinfo {author}
  {\bibfnamefont {M.~J.}\ \bibnamefont {Stevens}}, \bibinfo {author}
  {\bibfnamefont {T.}~\bibnamefont {Gerrits}}, \bibinfo {author} {\bibfnamefont
  {S.}~\bibnamefont {Glancy}}, \bibinfo {author} {\bibfnamefont {D.~R.}\
  \bibnamefont {Hamel}}, \bibinfo {author} {\bibfnamefont {M.~S.}\ \bibnamefont
  {Allman}}, \bibinfo {author} {\bibfnamefont {K.~J.}\ \bibnamefont {Coakley}},
  \bibinfo {author} {\bibfnamefont {S.~D.}\ \bibnamefont {Dyer}}, \bibinfo
  {author} {\bibfnamefont {C.}~\bibnamefont {Hodge}}, \bibinfo {author}
  {\bibfnamefont {A.~E.}\ \bibnamefont {Lita}}, \bibinfo {author}
  {\bibfnamefont {V.~B.}\ \bibnamefont {Verma}}, \bibinfo {author}
  {\bibfnamefont {C.}~\bibnamefont {Lambrocco}}, \bibinfo {author}
  {\bibfnamefont {E.}~\bibnamefont {Tortorici}}, \bibinfo {author}
  {\bibfnamefont {A.~L.}\ \bibnamefont {Migdall}}, \bibinfo {author}
  {\bibfnamefont {Y.}~\bibnamefont {Zhang}}, \bibinfo {author} {\bibfnamefont
  {D.~R.}\ \bibnamefont {Kumor}}, \bibinfo {author} {\bibfnamefont {W.~H.}\
  \bibnamefont {Farr}}, \bibinfo {author} {\bibfnamefont {F.}~\bibnamefont
  {Marsili}}, \bibinfo {author} {\bibfnamefont {M.~D.}\ \bibnamefont {Shaw}},
  \bibinfo {author} {\bibfnamefont {J.~A.}\ \bibnamefont {Stern}}, \bibinfo
  {author} {\bibfnamefont {C.}~\bibnamefont {Abell\'an}}, \bibinfo {author}
  {\bibfnamefont {W.}~\bibnamefont {Amaya}}, \bibinfo {author} {\bibfnamefont
  {V.}~\bibnamefont {Pruneri}}, \bibinfo {author} {\bibfnamefont
  {T.}~\bibnamefont {Jennewein}}, \bibinfo {author} {\bibfnamefont {M.~W.}\
  \bibnamefont {Mitchell}}, \bibinfo {author} {\bibfnamefont {P.~G.}\
  \bibnamefont {Kwiat}}, \bibinfo {author} {\bibfnamefont {J.~C.}\ \bibnamefont
  {Bienfang}}, \bibinfo {author} {\bibfnamefont {R.~P.}\ \bibnamefont {Mirin}},
  \bibinfo {author} {\bibfnamefont {E.}~\bibnamefont {Knill}},\ and\ \bibinfo
  {author} {\bibfnamefont {S.~W.}\ \bibnamefont {Nam}},\ }\href
  {https://doi.org/10.1103/PhysRevLett.115.250402} {\bibfield  {journal}
  {\bibinfo  {journal} {Phys. Rev. Lett.}\ }\textbf {\bibinfo {volume} {115}},\
  \bibinfo {pages} {250402} (\bibinfo {year} {2015})}\BibitemShut {NoStop}%
\bibitem [{\citenamefont {Vasilyev}\ \emph {et~al.}(2017)\citenamefont
  {Vasilyev}, \citenamefont {Schumann}, \citenamefont {Giebels}, \citenamefont
  {Gollisch}, \citenamefont {Kirschner},\ and\ \citenamefont
  {Feder}}]{2017Vasilyev}%
  \BibitemOpen
  \bibfield  {author} {\bibinfo {author} {\bibfnamefont {D.}~\bibnamefont
  {Vasilyev}}, \bibinfo {author} {\bibfnamefont {F.~O.}\ \bibnamefont
  {Schumann}}, \bibinfo {author} {\bibfnamefont {F.}~\bibnamefont {Giebels}},
  \bibinfo {author} {\bibfnamefont {H.}~\bibnamefont {Gollisch}}, \bibinfo
  {author} {\bibfnamefont {J.}~\bibnamefont {Kirschner}},\ and\ \bibinfo
  {author} {\bibfnamefont {R.}~\bibnamefont {Feder}},\ }\href
  {https://doi.org/10.1103/PhysRevB.95.115134} {\bibfield  {journal} {\bibinfo
  {journal} {Phys. Rev. B}\ }\textbf {\bibinfo {volume} {95}},\ \bibinfo
  {pages} {115134} (\bibinfo {year} {2017})}\BibitemShut {NoStop}%
\bibitem [{\citenamefont {Storz}\ \emph {et~al.}(2023)\citenamefont {Storz},
  \citenamefont {Sch\"{a}r}, \citenamefont {Kulikov}, \citenamefont {Magnard},
  \citenamefont {Kurpiers}, \citenamefont {L\"{u}tolf}, \citenamefont {Walter},
  \citenamefont {Copetudo}, \citenamefont {Reuer}, \citenamefont {Akin},
  \citenamefont {Besse}, \citenamefont {Gabureac}, \citenamefont {Norris},
  \citenamefont {Rosario}, \citenamefont {Martin}, \citenamefont {Martinez},
  \citenamefont {Amaya}, \citenamefont {Mitchell}, \citenamefont {Abellan},
  \citenamefont {Bancal}, \citenamefont {Sangouard}, \citenamefont {Royer},
  \citenamefont {Blais},\ and\ \citenamefont {Wallraff}}]{2023Storz}%
  \BibitemOpen
  \bibfield  {author} {\bibinfo {author} {\bibfnamefont {S.}~\bibnamefont
  {Storz}}, \bibinfo {author} {\bibfnamefont {J.}~\bibnamefont {Sch\"{a}r}},
  \bibinfo {author} {\bibfnamefont {A.}~\bibnamefont {Kulikov}}, \bibinfo
  {author} {\bibfnamefont {P.}~\bibnamefont {Magnard}}, \bibinfo {author}
  {\bibfnamefont {P.}~\bibnamefont {Kurpiers}}, \bibinfo {author}
  {\bibfnamefont {J.}~\bibnamefont {L\"{u}tolf}}, \bibinfo {author}
  {\bibfnamefont {T.}~\bibnamefont {Walter}}, \bibinfo {author} {\bibfnamefont
  {A.}~\bibnamefont {Copetudo}}, \bibinfo {author} {\bibfnamefont
  {K.}~\bibnamefont {Reuer}}, \bibinfo {author} {\bibfnamefont
  {A.}~\bibnamefont {Akin}}, \bibinfo {author} {\bibfnamefont {J.-C.}\
  \bibnamefont {Besse}}, \bibinfo {author} {\bibfnamefont {M.}~\bibnamefont
  {Gabureac}}, \bibinfo {author} {\bibfnamefont {G.~J.}\ \bibnamefont
  {Norris}}, \bibinfo {author} {\bibfnamefont {A.}~\bibnamefont {Rosario}},
  \bibinfo {author} {\bibfnamefont {F.}~\bibnamefont {Martin}}, \bibinfo
  {author} {\bibfnamefont {J.}~\bibnamefont {Martinez}}, \bibinfo {author}
  {\bibfnamefont {W.}~\bibnamefont {Amaya}}, \bibinfo {author} {\bibfnamefont
  {M.~W.}\ \bibnamefont {Mitchell}}, \bibinfo {author} {\bibfnamefont
  {C.}~\bibnamefont {Abellan}}, \bibinfo {author} {\bibfnamefont {J.-D.}\
  \bibnamefont {Bancal}}, \bibinfo {author} {\bibfnamefont {N.}~\bibnamefont
  {Sangouard}}, \bibinfo {author} {\bibfnamefont {B.}~\bibnamefont {Royer}},
  \bibinfo {author} {\bibfnamefont {A.}~\bibnamefont {Blais}},\ and\ \bibinfo
  {author} {\bibfnamefont {A.}~\bibnamefont {Wallraff}},\ }\href
  {https://doi.org/10.1038/s41586-023-05885-0} {\bibfield  {journal} {\bibinfo
  {journal} {Nature}\ }\textbf {\bibinfo {volume} {617}},\ \bibinfo {pages}
  {265} (\bibinfo {year} {2023})}\BibitemShut {NoStop}%
\bibitem [{\citenamefont {Barrett}\ \emph {et~al.}(2002)\citenamefont
  {Barrett}, \citenamefont {Collins}, \citenamefont {Hardy}, \citenamefont
  {Kent},\ and\ \citenamefont {Popescu}}]{2002Barrett}%
  \BibitemOpen
  \bibfield  {author} {\bibinfo {author} {\bibfnamefont {J.}~\bibnamefont
  {Barrett}}, \bibinfo {author} {\bibfnamefont {D.}~\bibnamefont {Collins}},
  \bibinfo {author} {\bibfnamefont {L.}~\bibnamefont {Hardy}}, \bibinfo
  {author} {\bibfnamefont {A.}~\bibnamefont {Kent}},\ and\ \bibinfo {author}
  {\bibfnamefont {S.}~\bibnamefont {Popescu}},\ }\href
  {https://doi.org/10.1103/PhysRevA.66.042111} {\bibfield  {journal} {\bibinfo
  {journal} {Phys. Rev. A}\ }\textbf {\bibinfo {volume} {66}},\ \bibinfo
  {pages} {042111} (\bibinfo {year} {2002})}\BibitemShut {NoStop}%
\bibitem [{\citenamefont {Lamehi-Rachti}\ and\ \citenamefont
  {Mittig}(1976)}]{1976Rachti}%
  \BibitemOpen
  \bibfield  {author} {\bibinfo {author} {\bibfnamefont {M.}~\bibnamefont
  {Lamehi-Rachti}}\ and\ \bibinfo {author} {\bibfnamefont {W.}~\bibnamefont
  {Mittig}},\ }\href {https://doi.org/10.1103/PhysRevD.14.2543} {\bibfield
  {journal} {\bibinfo  {journal} {Phys. Rev. D}\ }\textbf {\bibinfo {volume}
  {14}},\ \bibinfo {pages} {2543} (\bibinfo {year} {1976})}\BibitemShut
  {NoStop}%
\bibitem [{\citenamefont {Sakai}\ \emph {et~al.}(2006)\citenamefont {Sakai},
  \citenamefont {Saito}, \citenamefont {Ikeda}, \citenamefont {Itoh},
  \citenamefont {Kawabata}, \citenamefont {Kuboki}, \citenamefont {Maeda},
  \citenamefont {Matsui}, \citenamefont {Rangacharyulu}, \citenamefont
  {Sasano}, \citenamefont {Satou}, \citenamefont {Sekiguchi}, \citenamefont
  {Suda}, \citenamefont {Tamii}, \citenamefont {Uesaka},\ and\ \citenamefont
  {Yako}}]{2003Sakai}%
  \BibitemOpen
  \bibfield  {author} {\bibinfo {author} {\bibfnamefont {H.}~\bibnamefont
  {Sakai}}, \bibinfo {author} {\bibfnamefont {T.}~\bibnamefont {Saito}},
  \bibinfo {author} {\bibfnamefont {T.}~\bibnamefont {Ikeda}}, \bibinfo
  {author} {\bibfnamefont {K.}~\bibnamefont {Itoh}}, \bibinfo {author}
  {\bibfnamefont {T.}~\bibnamefont {Kawabata}}, \bibinfo {author}
  {\bibfnamefont {H.}~\bibnamefont {Kuboki}}, \bibinfo {author} {\bibfnamefont
  {Y.}~\bibnamefont {Maeda}}, \bibinfo {author} {\bibfnamefont
  {N.}~\bibnamefont {Matsui}}, \bibinfo {author} {\bibfnamefont
  {C.}~\bibnamefont {Rangacharyulu}}, \bibinfo {author} {\bibfnamefont
  {M.}~\bibnamefont {Sasano}}, \bibinfo {author} {\bibfnamefont
  {Y.}~\bibnamefont {Satou}}, \bibinfo {author} {\bibfnamefont
  {K.}~\bibnamefont {Sekiguchi}}, \bibinfo {author} {\bibfnamefont
  {K.}~\bibnamefont {Suda}}, \bibinfo {author} {\bibfnamefont {A.}~\bibnamefont
  {Tamii}}, \bibinfo {author} {\bibfnamefont {T.}~\bibnamefont {Uesaka}},\ and\
  \bibinfo {author} {\bibfnamefont {K.}~\bibnamefont {Yako}},\ }\href
  {https://doi.org/10.1103/PhysRevLett.97.150405} {\bibfield  {journal}
  {\bibinfo  {journal} {Phys. Rev. Lett.}\ }\textbf {\bibinfo {volume} {97}},\
  \bibinfo {pages} {150405} (\bibinfo {year} {2006})}\BibitemShut {NoStop}%
\bibitem [{\citenamefont {Kanada-En’yo}(2015)}]{2015Kanada}%
  \BibitemOpen
  \bibfield  {author} {\bibinfo {author} {\bibfnamefont {Y.}~\bibnamefont
  {Kanada-En’yo}},\ }\href {https://doi.org/10.1093/ptep/ptv050} {\bibfield
  {journal} {\bibinfo  {journal} {Progress of Theoretical and Experimental
  Physics}\ }\textbf {\bibinfo {volume} {2015}},\ \bibinfo {pages} {043D04}
  (\bibinfo {year} {2015})}\BibitemShut {NoStop}%
\bibitem [{\citenamefont {Bulgac}(2023)}]{2023Bulgac_PRC}%
  \BibitemOpen
  \bibfield  {author} {\bibinfo {author} {\bibfnamefont {A.}~\bibnamefont
  {Bulgac}},\ }\href {https://doi.org/10.1103/PhysRevC.107.L061602} {\bibfield
  {journal} {\bibinfo  {journal} {Phys. Rev. C}\ }\textbf {\bibinfo {volume}
  {107}},\ \bibinfo {pages} {L061602} (\bibinfo {year} {2023})}\BibitemShut
  {NoStop}%
\bibitem [{\citenamefont {Pazy}(2023)}]{2023Pazy}%
  \BibitemOpen
  \bibfield  {author} {\bibinfo {author} {\bibfnamefont {E.}~\bibnamefont
  {Pazy}},\ }\href {https://doi.org/10.1103/PhysRevC.107.054308} {\bibfield
  {journal} {\bibinfo  {journal} {Phys. Rev. C}\ }\textbf {\bibinfo {volume}
  {107}},\ \bibinfo {pages} {054308} (\bibinfo {year} {2023})}\BibitemShut
  {NoStop}%
\bibitem [{\citenamefont {Gu}\ \emph {et~al.}(2023)\citenamefont {Gu},
  \citenamefont {Sun}, \citenamefont {Hagen},\ and\ \citenamefont
  {Papenbrock}}]{2023Gu}%
  \BibitemOpen
  \bibfield  {author} {\bibinfo {author} {\bibfnamefont {C.}~\bibnamefont
  {Gu}}, \bibinfo {author} {\bibfnamefont {Z.~H.}\ \bibnamefont {Sun}},
  \bibinfo {author} {\bibfnamefont {G.}~\bibnamefont {Hagen}},\ and\ \bibinfo
  {author} {\bibfnamefont {T.}~\bibnamefont {Papenbrock}},\ }\href
  {https://doi.org/10.1103/PhysRevC.108.054309} {\bibfield  {journal} {\bibinfo
   {journal} {Phys. Rev. C}\ }\textbf {\bibinfo {volume} {108}},\ \bibinfo
  {pages} {054309} (\bibinfo {year} {2023})}\BibitemShut {NoStop}%
\bibitem [{\citenamefont {Hengstenberg}\ \emph {et~al.}(2023)\citenamefont
  {Hengstenberg}, \citenamefont {Robin},\ and\ \citenamefont
  {Savage}}]{2023Heng}%
  \BibitemOpen
  \bibfield  {author} {\bibinfo {author} {\bibfnamefont {S.~M.}\ \bibnamefont
  {Hengstenberg}}, \bibinfo {author} {\bibfnamefont {C.~E.~P.}\ \bibnamefont
  {Robin}},\ and\ \bibinfo {author} {\bibfnamefont {M.~J.}\ \bibnamefont
  {Savage}},\ }\href {https://doi.org/10.1140/epja/s10050-023-01145-x}
  {\bibfield  {journal} {\bibinfo  {journal} {The European Physical Journal A}\
  }\textbf {\bibinfo {volume} {59}},\ \bibinfo {pages} {231} (\bibinfo {year}
  {2023})}\BibitemShut {NoStop}%
\bibitem [{\citenamefont {P\'{e}rez-Obiol}\ \emph {et~al.}(2023)\citenamefont
  {P\'{e}rez-Obiol}, \citenamefont {Masot-Llima}, \citenamefont {Romero},
  \citenamefont {Men\'{e}ndez}, \citenamefont {Rios}, \citenamefont
  {Garcia-S\'{a}ez},\ and\ \citenamefont {Juli\'{a}-Diaz}}]{2023Obiol}%
  \BibitemOpen
  \bibfield  {author} {\bibinfo {author} {\bibfnamefont {A.}~\bibnamefont
  {P\'{e}rez-Obiol}}, \bibinfo {author} {\bibfnamefont {S.}~\bibnamefont
  {Masot-Llima}}, \bibinfo {author} {\bibfnamefont {A.~M.}\ \bibnamefont
  {Romero}}, \bibinfo {author} {\bibfnamefont {J.}~\bibnamefont
  {Men\'{e}ndez}}, \bibinfo {author} {\bibfnamefont {A.}~\bibnamefont {Rios}},
  \bibinfo {author} {\bibfnamefont {A.}~\bibnamefont {Garcia-S\'{a}ez}},\ and\
  \bibinfo {author} {\bibfnamefont {B.}~\bibnamefont {Juli\'{a}-Diaz}},\ }\href
  {https://doi.org/10.1140/epja/s10050-023-01151-z} {\bibfield  {journal}
  {\bibinfo  {journal} {The European Physical Journal A}\ }\textbf {\bibinfo
  {volume} {59}},\ \bibinfo {pages} {240} (\bibinfo {year} {2023})}\BibitemShut
  {NoStop}%
\bibitem [{\citenamefont {Kou}\ \emph {et~al.}(2024)\citenamefont {Kou},
  \citenamefont {Chen},\ and\ \citenamefont {Chen}}]{2024Kou}%
  \BibitemOpen
  \bibfield  {author} {\bibinfo {author} {\bibfnamefont {W.}~\bibnamefont
  {Kou}}, \bibinfo {author} {\bibfnamefont {J.}~\bibnamefont {Chen}},\ and\
  \bibinfo {author} {\bibfnamefont {X.}~\bibnamefont {Chen}},\ }\href
  {https://doi.org/https://doi.org/10.1016/j.physletb.2024.138453} {\bibfield
  {journal} {\bibinfo  {journal} {Physics Letters B}\ }\textbf {\bibinfo
  {volume} {849}},\ \bibinfo {pages} {138453} (\bibinfo {year}
  {2024})}\BibitemShut {NoStop}%
\bibitem [{\citenamefont {Kirchner}\ \emph {et~al.}(2024)\citenamefont
  {Kirchner}, \citenamefont {Elkamhawy},\ and\ \citenamefont
  {Hammer}}]{2024Krichner}%
  \BibitemOpen
  \bibfield  {author} {\bibinfo {author} {\bibfnamefont {T.}~\bibnamefont
  {Kirchner}}, \bibinfo {author} {\bibfnamefont {W.}~\bibnamefont
  {Elkamhawy}},\ and\ \bibinfo {author} {\bibfnamefont {H.-W.}\ \bibnamefont
  {Hammer}},\ }\href {https://doi.org/10.1007/s00601-024-01897-2} {\bibfield
  {journal} {\bibinfo  {journal} {Few-Body Systems}\ }\textbf {\bibinfo
  {volume} {65}},\ \bibinfo {pages} {29} (\bibinfo {year} {2024})}\BibitemShut
  {NoStop}%
\bibitem [{\citenamefont {Robin}\ \emph {et~al.}(2021)\citenamefont {Robin},
  \citenamefont {Savage},\ and\ \citenamefont {Pillet}}]{2021Robin}%
  \BibitemOpen
  \bibfield  {author} {\bibinfo {author} {\bibfnamefont {C.}~\bibnamefont
  {Robin}}, \bibinfo {author} {\bibfnamefont {M.~J.}\ \bibnamefont {Savage}},\
  and\ \bibinfo {author} {\bibfnamefont {N.}~\bibnamefont {Pillet}},\ }\href
  {https://doi.org/10.1103/PhysRevC.103.034325} {\bibfield  {journal} {\bibinfo
   {journal} {Phys. Rev. C}\ }\textbf {\bibinfo {volume} {103}},\ \bibinfo
  {pages} {034325} (\bibinfo {year} {2021})}\BibitemShut {NoStop}%
\bibitem [{\citenamefont {Kruppa}\ \emph {et~al.}(2021)\citenamefont {Kruppa},
  \citenamefont {Kov\'{a}cs}, \citenamefont {Salamon},\ and\ \citenamefont
  {Legeza}}]{2021Kruppa}%
  \BibitemOpen
  \bibfield  {author} {\bibinfo {author} {\bibfnamefont {A.~T.}\ \bibnamefont
  {Kruppa}}, \bibinfo {author} {\bibfnamefont {J.}~\bibnamefont {Kov\'{a}cs}},
  \bibinfo {author} {\bibfnamefont {P.}~\bibnamefont {Salamon}},\ and\ \bibinfo
  {author} {\bibfnamefont {O.}~\bibnamefont {Legeza}},\ }\href
  {https://doi.org/10.1088/1361-6471/abc2dd} {\bibfield  {journal} {\bibinfo
  {journal} {Journal of Physics G: Nuclear and Particle Physics}\ }\textbf
  {\bibinfo {volume} {48}},\ \bibinfo {pages} {025107} (\bibinfo {year}
  {2021})}\BibitemShut {NoStop}%
\bibitem [{\citenamefont {Miller}(2023{\natexlab{a}})}]{2023Miller}%
  \BibitemOpen
  \bibfield  {author} {\bibinfo {author} {\bibfnamefont {G.~A.}\ \bibnamefont
  {Miller}},\ }\href {https://doi.org/10.1103/PhysRevC.108.L041601} {\bibfield
  {journal} {\bibinfo  {journal} {Phys. Rev. C}\ }\textbf {\bibinfo {volume}
  {108}},\ \bibinfo {pages} {L041601} (\bibinfo {year}
  {2023}{\natexlab{a}})}\BibitemShut {NoStop}%
\bibitem [{\citenamefont {Miller}(2023{\natexlab{b}})}]{2023Miller_PRCLett}%
  \BibitemOpen
  \bibfield  {author} {\bibinfo {author} {\bibfnamefont {G.~A.}\ \bibnamefont
  {Miller}},\ }\href {https://doi.org/10.1103/PhysRevC.108.L031002} {\bibfield
  {journal} {\bibinfo  {journal} {Phys. Rev. C}\ }\textbf {\bibinfo {volume}
  {108}},\ \bibinfo {pages} {L031002} (\bibinfo {year}
  {2023}{\natexlab{b}})}\BibitemShut {NoStop}%
\bibitem [{\citenamefont {Tichai}\ \emph {et~al.}(2023)\citenamefont {Tichai},
  \citenamefont {Knecht}, \citenamefont {Kruppa}, \citenamefont {Legeza},
  \citenamefont {Moca}, \citenamefont {Schwenk}, \citenamefont {Werner},\ and\
  \citenamefont {Zarand}}]{2023Tichai}%
  \BibitemOpen
  \bibfield  {author} {\bibinfo {author} {\bibfnamefont {A.}~\bibnamefont
  {Tichai}}, \bibinfo {author} {\bibfnamefont {S.}~\bibnamefont {Knecht}},
  \bibinfo {author} {\bibfnamefont {A.}~\bibnamefont {Kruppa}}, \bibinfo
  {author} {\bibfnamefont {O.}~\bibnamefont {Legeza}}, \bibinfo {author}
  {\bibfnamefont {C.}~\bibnamefont {Moca}}, \bibinfo {author} {\bibfnamefont
  {A.}~\bibnamefont {Schwenk}}, \bibinfo {author} {\bibfnamefont
  {M.}~\bibnamefont {Werner}},\ and\ \bibinfo {author} {\bibfnamefont
  {G.}~\bibnamefont {Zarand}},\ }\href
  {https://doi.org/https://doi.org/10.1016/j.physletb.2023.138139} {\bibfield
  {journal} {\bibinfo  {journal} {Physics Letters B}\ }\textbf {\bibinfo
  {volume} {845}},\ \bibinfo {pages} {138139} (\bibinfo {year}
  {2023})}\BibitemShut {NoStop}%
\bibitem [{\citenamefont {Sun}\ \emph {et~al.}(2023)\citenamefont {Sun},
  \citenamefont {Hagen},\ and\ \citenamefont {Papenbrock}}]{2023Sun_PRC}%
  \BibitemOpen
  \bibfield  {author} {\bibinfo {author} {\bibfnamefont {Z.~H.}\ \bibnamefont
  {Sun}}, \bibinfo {author} {\bibfnamefont {G.}~\bibnamefont {Hagen}},\ and\
  \bibinfo {author} {\bibfnamefont {T.}~\bibnamefont {Papenbrock}},\ }\href
  {https://doi.org/10.1103/PhysRevC.108.014307} {\bibfield  {journal} {\bibinfo
   {journal} {Phys. Rev. C}\ }\textbf {\bibinfo {volume} {108}},\ \bibinfo
  {pages} {014307} (\bibinfo {year} {2023})}\BibitemShut {NoStop}%
\bibitem [{\citenamefont {Johnson}\ and\ \citenamefont
  {Gorton}(2023)}]{2023Johnson}%
  \BibitemOpen
  \bibfield  {author} {\bibinfo {author} {\bibfnamefont {C.~W.}\ \bibnamefont
  {Johnson}}\ and\ \bibinfo {author} {\bibfnamefont {O.~C.}\ \bibnamefont
  {Gorton}},\ }\href {https://doi.org/10.1088/1361-6471/acbece} {\bibfield
  {journal} {\bibinfo  {journal} {Journal of Physics G: Nuclear and Particle
  Physics}\ }\textbf {\bibinfo {volume} {50}},\ \bibinfo {pages} {045110}
  (\bibinfo {year} {2023})}\BibitemShut {NoStop}%
\bibitem [{\citenamefont {Bai}(2024)}]{2024Bai}%
  \BibitemOpen
  \bibfield  {author} {\bibinfo {author} {\bibfnamefont {D.}~\bibnamefont
  {Bai}},\ }\href {https://doi.org/10.1103/PhysRevC.109.034001} {\bibfield
  {journal} {\bibinfo  {journal} {Phys. Rev. C}\ }\textbf {\bibinfo {volume}
  {109}},\ \bibinfo {pages} {034001} (\bibinfo {year} {2024})}\BibitemShut
  {NoStop}%
\bibitem [{\citenamefont {Bertulani}\ \emph {et~al.}(2008)\citenamefont
  {Bertulani}, \citenamefont {Hussein},\ and\ \citenamefont
  {Verde}}]{08Bertulani}%
  \BibitemOpen
  \bibfield  {author} {\bibinfo {author} {\bibfnamefont {C.}~\bibnamefont
  {Bertulani}}, \bibinfo {author} {\bibfnamefont {M.}~\bibnamefont {Hussein}},\
  and\ \bibinfo {author} {\bibfnamefont {G.}~\bibnamefont {Verde}},\ }\href
  {https://doi.org/http://dx.doi.org/10.1016/j.physletb.2008.06.062} {\bibfield
   {journal} {\bibinfo  {journal} {Physics Letters B}\ }\textbf {\bibinfo
  {volume} {666}},\ \bibinfo {pages} {86 } (\bibinfo {year}
  {2008})}\BibitemShut {NoStop}%
\bibitem [{\citenamefont {Grigorenko}(2009)}]{2009Gri_rev}%
  \BibitemOpen
  \bibfield  {author} {\bibinfo {author} {\bibfnamefont {L.~V.}\ \bibnamefont
  {Grigorenko}},\ }\href {https://doi.org/10.1134/S1063779609050049} {\bibfield
   {journal} {\bibinfo  {journal} {Physics of Particles and Nuclei}\ }\textbf
  {\bibinfo {volume} {40}},\ \bibinfo {pages} {674} (\bibinfo {year}
  {2009})}\BibitemShut {NoStop}%
\bibitem [{\citenamefont {Pf\"utzner}\ \emph {et~al.}(2012)\citenamefont
  {Pf\"utzner}, \citenamefont {Karny}, \citenamefont {Grigorenko},\ and\
  \citenamefont {Riisager}}]{2012Pfu_rev}%
  \BibitemOpen
  \bibfield  {author} {\bibinfo {author} {\bibfnamefont {M.}~\bibnamefont
  {Pf\"utzner}}, \bibinfo {author} {\bibfnamefont {M.}~\bibnamefont {Karny}},
  \bibinfo {author} {\bibfnamefont {L.~V.}\ \bibnamefont {Grigorenko}},\ and\
  \bibinfo {author} {\bibfnamefont {K.}~\bibnamefont {Riisager}},\ }\href
  {https://doi.org/10.1103/RevModPhys.84.567} {\bibfield  {journal} {\bibinfo
  {journal} {Rev. Mod. Phys.}\ }\textbf {\bibinfo {volume} {84}},\ \bibinfo
  {pages} {567} (\bibinfo {year} {2012})}\BibitemShut {NoStop}%
\bibitem [{\citenamefont {Blank}\ and\ \citenamefont
  {Ploszajczak}(2008)}]{08Blank_01}%
  \BibitemOpen
  \bibfield  {author} {\bibinfo {author} {\bibfnamefont {B.}~\bibnamefont
  {Blank}}\ and\ \bibinfo {author} {\bibfnamefont {M.}~\bibnamefont
  {Ploszajczak}},\ }\href {http://stacks.iop.org/0034-4885/71/i=4/a=046301}
  {\bibfield  {journal} {\bibinfo  {journal} {Reports on Progress in Physics}\
  }\textbf {\bibinfo {volume} {71}},\ \bibinfo {pages} {046301} (\bibinfo
  {year} {2008})}\BibitemShut {NoStop}%
\bibitem [{\citenamefont {Blank}\ and\ \citenamefont
  {Borge}(2008)}]{08Blank_02}%
  \BibitemOpen
  \bibfield  {author} {\bibinfo {author} {\bibfnamefont {B.}~\bibnamefont
  {Blank}}\ and\ \bibinfo {author} {\bibfnamefont {M.}~\bibnamefont {Borge}},\
  }\href {https://doi.org/https://doi.org/10.1016/j.ppnp.2007.12.001}
  {\bibfield  {journal} {\bibinfo  {journal} {Progress in Particle and Nuclear
  Physics}\ }\textbf {\bibinfo {volume} {60}},\ \bibinfo {pages} {403 }
  (\bibinfo {year} {2008})}\BibitemShut {NoStop}%
\bibitem [{\citenamefont {Pf\"{u}tzner}\ \emph {et~al.}(2023)\citenamefont
  {Pf\"{u}tzner}, \citenamefont {Mukha},\ and\ \citenamefont
  {Wang}}]{2023Pfutzner_rev}%
  \BibitemOpen
  \bibfield  {author} {\bibinfo {author} {\bibfnamefont {M.}~\bibnamefont
  {Pf\"{u}tzner}}, \bibinfo {author} {\bibfnamefont {I.}~\bibnamefont
  {Mukha}},\ and\ \bibinfo {author} {\bibfnamefont {S.}~\bibnamefont {Wang}},\
  }\href {https://doi.org/https://doi.org/10.1016/j.ppnp.2023.104050}
  {\bibfield  {journal} {\bibinfo  {journal} {Progress in Particle and Nuclear
  Physics}\ }\textbf {\bibinfo {volume} {132}},\ \bibinfo {pages} {104050}
  (\bibinfo {year} {2023})}\BibitemShut {NoStop}%
\bibitem [{\citenamefont {Qi}\ \emph {et~al.}(2019)\citenamefont {Qi},
  \citenamefont {Liotta},\ and\ \citenamefont {Wyss}}]{2019Qi_rev}%
  \BibitemOpen
  \bibfield  {author} {\bibinfo {author} {\bibfnamefont {C.}~\bibnamefont
  {Qi}}, \bibinfo {author} {\bibfnamefont {R.}~\bibnamefont {Liotta}},\ and\
  \bibinfo {author} {\bibfnamefont {R.}~\bibnamefont {Wyss}},\ }\href
  {https://doi.org/https://doi.org/10.1016/j.ppnp.2018.11.003} {\bibfield
  {journal} {\bibinfo  {journal} {Progress in Particle and Nuclear Physics}\
  }\textbf {\bibinfo {volume} {105}},\ \bibinfo {pages} {214} (\bibinfo {year}
  {2019})}\BibitemShut {NoStop}%
\bibitem [{\citenamefont {Goldansky}(1960)}]{60Gold}%
  \BibitemOpen
  \bibfield  {author} {\bibinfo {author} {\bibfnamefont {V.~I.}\ \bibnamefont
  {Goldansky}},\ }\href@noop {} {\bibfield  {journal} {\bibinfo  {journal}
  {Nucl. Phys.}\ }\textbf {\bibinfo {volume} {19}},\ \bibinfo {pages} {482}
  (\bibinfo {year} {1960})}\BibitemShut {NoStop}%
\bibitem [{\citenamefont {Goldansky}(1961)}]{61Gold}%
  \BibitemOpen
  \bibfield  {author} {\bibinfo {author} {\bibfnamefont {V.~I.}\ \bibnamefont
  {Goldansky}},\ }\href@noop {} {\bibfield  {journal} {\bibinfo  {journal}
  {Nucl. Phys.}\ }\textbf {\bibinfo {volume} {27}},\ \bibinfo {pages} {648}
  (\bibinfo {year} {1961})}\BibitemShut {NoStop}%
\bibitem [{\citenamefont {Benatti}\ \emph {et~al.}(2020)\citenamefont
  {Benatti}, \citenamefont {Floreanini}, \citenamefont {Franchini},\ and\
  \citenamefont {Marzolino}}]{2020Benatti}%
  \BibitemOpen
  \bibfield  {author} {\bibinfo {author} {\bibfnamefont {F.}~\bibnamefont
  {Benatti}}, \bibinfo {author} {\bibfnamefont {R.}~\bibnamefont {Floreanini}},
  \bibinfo {author} {\bibfnamefont {F.}~\bibnamefont {Franchini}},\ and\
  \bibinfo {author} {\bibfnamefont {U.}~\bibnamefont {Marzolino}},\ }\href
  {https://doi.org/https://doi.org/10.1016/j.physrep.2020.07.003} {\bibfield
  {journal} {\bibinfo  {journal} {Physics Reports}\ }\textbf {\bibinfo {volume}
  {878}},\ \bibinfo {pages} {1} (\bibinfo {year} {2020})}\BibitemShut {NoStop}%
\bibitem [{\citenamefont {Johann}\ and\ \citenamefont
  {Marzolino}(2021)}]{2021Frederick}%
  \BibitemOpen
  \bibfield  {author} {\bibinfo {author} {\bibfnamefont {T.~J.~F.}\
  \bibnamefont {Johann}}\ and\ \bibinfo {author} {\bibfnamefont
  {U.}~\bibnamefont {Marzolino}},\ }\href
  {https://doi.org/10.1038/s41598-021-94991-y} {\bibfield  {journal} {\bibinfo
  {journal} {Scientific Reports}\ }\textbf {\bibinfo {volume} {11}},\ \bibinfo
  {pages} {15478} (\bibinfo {year} {2021})}\BibitemShut {NoStop}%
\bibitem [{\citenamefont {Pa\ifmmode~\check{s}\else \v{s}\fi{}kauskas}\ and\
  \citenamefont {You}(2001)}]{2001Paskauskas}%
  \BibitemOpen
  \bibfield  {author} {\bibinfo {author} {\bibfnamefont {R.}~\bibnamefont
  {Pa\ifmmode~\check{s}\else \v{s}\fi{}kauskas}}\ and\ \bibinfo {author}
  {\bibfnamefont {L.}~\bibnamefont {You}},\ }\href
  {https://doi.org/10.1103/PhysRevA.64.042310} {\bibfield  {journal} {\bibinfo
  {journal} {Phys. Rev. A}\ }\textbf {\bibinfo {volume} {64}},\ \bibinfo
  {pages} {042310} (\bibinfo {year} {2001})}\BibitemShut {NoStop}%
\bibitem [{\citenamefont {Eckert}\ \emph {et~al.}(2002)\citenamefont {Eckert},
  \citenamefont {Schliemann}, \citenamefont {Bru{\ss}},\ and\ \citenamefont
  {Lewenstein}}]{2002Eckert}%
  \BibitemOpen
  \bibfield  {author} {\bibinfo {author} {\bibfnamefont {K.}~\bibnamefont
  {Eckert}}, \bibinfo {author} {\bibfnamefont {J.}~\bibnamefont {Schliemann}},
  \bibinfo {author} {\bibfnamefont {D.}~\bibnamefont {Bru{\ss}}},\ and\
  \bibinfo {author} {\bibfnamefont {M.}~\bibnamefont {Lewenstein}},\ }\href
  {https://doi.org/https://doi.org/10.1006/aphy.2002.6268} {\bibfield
  {journal} {\bibinfo  {journal} {Annals of Physics}\ }\textbf {\bibinfo
  {volume} {299}},\ \bibinfo {pages} {88} (\bibinfo {year} {2002})}\BibitemShut
  {NoStop}%
\bibitem [{\citenamefont {Ghirardi}\ \emph {et~al.}(2002)\citenamefont
  {Ghirardi}, \citenamefont {Marinatto},\ and\ \citenamefont
  {Weber}}]{2002Ghirardi}%
  \BibitemOpen
  \bibfield  {author} {\bibinfo {author} {\bibfnamefont {G.-C.}\ \bibnamefont
  {Ghirardi}}, \bibinfo {author} {\bibfnamefont {L.}~\bibnamefont
  {Marinatto}},\ and\ \bibinfo {author} {\bibfnamefont {T.}~\bibnamefont
  {Weber}},\ }\href {https://doi.org/10.1023/A:1015439502289} {\bibfield
  {journal} {\bibinfo  {journal} {Journal of Statistical Physics}\ }\textbf
  {\bibinfo {volume} {108}},\ \bibinfo {pages} {49} (\bibinfo {year}
  {2002})}\BibitemShut {NoStop}%
\bibitem [{\citenamefont {Ghirardi}\ and\ \citenamefont
  {Marinatto}(2004)}]{2004Ghirardi}%
  \BibitemOpen
  \bibfield  {author} {\bibinfo {author} {\bibfnamefont {G.}~\bibnamefont
  {Ghirardi}}\ and\ \bibinfo {author} {\bibfnamefont {L.}~\bibnamefont
  {Marinatto}},\ }\href {https://doi.org/10.1103/PhysRevA.70.012109} {\bibfield
   {journal} {\bibinfo  {journal} {Phys. Rev. A}\ }\textbf {\bibinfo {volume}
  {70}},\ \bibinfo {pages} {012109} (\bibinfo {year} {2004})}\BibitemShut
  {NoStop}%
\bibitem [{\citenamefont {Shi}(2003)}]{2003Shi}%
  \BibitemOpen
  \bibfield  {author} {\bibinfo {author} {\bibfnamefont {Y.}~\bibnamefont
  {Shi}},\ }\href {https://doi.org/10.1103/PhysRevA.67.024301} {\bibfield
  {journal} {\bibinfo  {journal} {Phys. Rev. A}\ }\textbf {\bibinfo {volume}
  {67}},\ \bibinfo {pages} {024301} (\bibinfo {year} {2003})}\BibitemShut
  {NoStop}%
\bibitem [{\citenamefont {Barnum}\ \emph {et~al.}(2004)\citenamefont {Barnum},
  \citenamefont {Knill}, \citenamefont {Ortiz}, \citenamefont {Somma},\ and\
  \citenamefont {Viola}}]{2004Barnum}%
  \BibitemOpen
  \bibfield  {author} {\bibinfo {author} {\bibfnamefont {H.}~\bibnamefont
  {Barnum}}, \bibinfo {author} {\bibfnamefont {E.}~\bibnamefont {Knill}},
  \bibinfo {author} {\bibfnamefont {G.}~\bibnamefont {Ortiz}}, \bibinfo
  {author} {\bibfnamefont {R.}~\bibnamefont {Somma}},\ and\ \bibinfo {author}
  {\bibfnamefont {L.}~\bibnamefont {Viola}},\ }\href
  {https://doi.org/10.1103/PhysRevLett.92.107902} {\bibfield  {journal}
  {\bibinfo  {journal} {Phys. Rev. Lett.}\ }\textbf {\bibinfo {volume} {92}},\
  \bibinfo {pages} {107902} (\bibinfo {year} {2004})}\BibitemShut {NoStop}%
\bibitem [{\citenamefont {Zanardi}\ \emph {et~al.}(2004)\citenamefont
  {Zanardi}, \citenamefont {Lidar},\ and\ \citenamefont {Lloyd}}]{2004Zanardi}%
  \BibitemOpen
  \bibfield  {author} {\bibinfo {author} {\bibfnamefont {P.}~\bibnamefont
  {Zanardi}}, \bibinfo {author} {\bibfnamefont {D.~A.}\ \bibnamefont {Lidar}},\
  and\ \bibinfo {author} {\bibfnamefont {S.}~\bibnamefont {Lloyd}},\ }\href
  {https://doi.org/10.1103/PhysRevLett.92.060402} {\bibfield  {journal}
  {\bibinfo  {journal} {Phys. Rev. Lett.}\ }\textbf {\bibinfo {volume} {92}},\
  \bibinfo {pages} {060402} (\bibinfo {year} {2004})}\BibitemShut {NoStop}%
\bibitem [{\citenamefont {Sasaki}\ \emph {et~al.}(2011)\citenamefont {Sasaki},
  \citenamefont {Ichikawa},\ and\ \citenamefont {Tsutsui}}]{2011Sasaki}%
  \BibitemOpen
  \bibfield  {author} {\bibinfo {author} {\bibfnamefont {T.}~\bibnamefont
  {Sasaki}}, \bibinfo {author} {\bibfnamefont {T.}~\bibnamefont {Ichikawa}},\
  and\ \bibinfo {author} {\bibfnamefont {I.}~\bibnamefont {Tsutsui}},\ }\href
  {https://doi.org/10.1103/PhysRevA.83.012113} {\bibfield  {journal} {\bibinfo
  {journal} {Phys. Rev. A}\ }\textbf {\bibinfo {volume} {83}},\ \bibinfo
  {pages} {012113} (\bibinfo {year} {2011})}\BibitemShut {NoStop}%
\bibitem [{\citenamefont {Balachandran}\ \emph {et~al.}(2013)\citenamefont
  {Balachandran}, \citenamefont {Govindarajan}, \citenamefont {de~Queiroz},\
  and\ \citenamefont {Reyes-Lega}}]{2013Balacha}%
  \BibitemOpen
  \bibfield  {author} {\bibinfo {author} {\bibfnamefont {A.~P.}\ \bibnamefont
  {Balachandran}}, \bibinfo {author} {\bibfnamefont {T.~R.}\ \bibnamefont
  {Govindarajan}}, \bibinfo {author} {\bibfnamefont {A.~R.}\ \bibnamefont
  {de~Queiroz}},\ and\ \bibinfo {author} {\bibfnamefont {A.~F.}\ \bibnamefont
  {Reyes-Lega}},\ }\href {https://doi.org/10.1103/PhysRevLett.110.080503}
  {\bibfield  {journal} {\bibinfo  {journal} {Phys. Rev. Lett.}\ }\textbf
  {\bibinfo {volume} {110}},\ \bibinfo {pages} {080503} (\bibinfo {year}
  {2013})}\BibitemShut {NoStop}%
\bibitem [{\citenamefont {Reusch}\ \emph {et~al.}(2015)\citenamefont {Reusch},
  \citenamefont {Sperling},\ and\ \citenamefont {Vogel}}]{2015Reusch}%
  \BibitemOpen
  \bibfield  {author} {\bibinfo {author} {\bibfnamefont {A.}~\bibnamefont
  {Reusch}}, \bibinfo {author} {\bibfnamefont {J.}~\bibnamefont {Sperling}},\
  and\ \bibinfo {author} {\bibfnamefont {W.}~\bibnamefont {Vogel}},\ }\href
  {https://doi.org/10.1103/PhysRevA.91.042324} {\bibfield  {journal} {\bibinfo
  {journal} {Phys. Rev. A}\ }\textbf {\bibinfo {volume} {91}},\ \bibinfo
  {pages} {042324} (\bibinfo {year} {2015})}\BibitemShut {NoStop}%
\bibitem [{\citenamefont {Tichy}\ \emph {et~al.}(2011)\citenamefont {Tichy},
  \citenamefont {Mintert},\ and\ \citenamefont {Buchleitner}}]{2011Tichy}%
  \BibitemOpen
  \bibfield  {author} {\bibinfo {author} {\bibfnamefont {M.~C.}\ \bibnamefont
  {Tichy}}, \bibinfo {author} {\bibfnamefont {F.}~\bibnamefont {Mintert}},\
  and\ \bibinfo {author} {\bibfnamefont {A.}~\bibnamefont {Buchleitner}},\
  }\href {https://doi.org/10.1088/0953-4075/44/19/192001} {\bibfield  {journal}
  {\bibinfo  {journal} {Journal of Physics B: Atomic, Molecular and Optical
  Physics}\ }\textbf {\bibinfo {volume} {44}},\ \bibinfo {pages} {192001}
  (\bibinfo {year} {2011})},\ \bibinfo {note} {corrections were made to this
  article on 28 September 2011, the received date was incorrectly
  given.}\BibitemShut {Stop}%
\bibitem [{\citenamefont {Lo~Franco}\ and\ \citenamefont
  {Compagno}(2016)}]{2016Franco}%
  \BibitemOpen
  \bibfield  {author} {\bibinfo {author} {\bibfnamefont {R.}~\bibnamefont
  {Lo~Franco}}\ and\ \bibinfo {author} {\bibfnamefont {G.}~\bibnamefont
  {Compagno}},\ }\href {https://doi.org/10.1038/srep20603} {\bibfield
  {journal} {\bibinfo  {journal} {Scientific Reports}\ }\textbf {\bibinfo
  {volume} {6}},\ \bibinfo {pages} {20603} (\bibinfo {year}
  {2016})}\BibitemShut {NoStop}%
\bibitem [{\citenamefont {Benatti}\ \emph {et~al.}(2017)\citenamefont
  {Benatti}, \citenamefont {Floreanini}, \citenamefont {Franchini},\ and\
  \citenamefont {Marzolino}}]{2017Benatti}%
  \BibitemOpen
  \bibfield  {author} {\bibinfo {author} {\bibfnamefont {F.}~\bibnamefont
  {Benatti}}, \bibinfo {author} {\bibfnamefont {R.}~\bibnamefont {Floreanini}},
  \bibinfo {author} {\bibfnamefont {F.}~\bibnamefont {Franchini}},\ and\
  \bibinfo {author} {\bibfnamefont {U.}~\bibnamefont {Marzolino}},\ }\href
  {https://doi.org/10.1142/S1230161217400042} {\bibfield  {journal} {\bibinfo
  {journal} {Open Systems \& Information Dynamics}\ }\textbf {\bibinfo {volume}
  {24}},\ \bibinfo {pages} {1740004} (\bibinfo {year} {2017})}\BibitemShut
  {NoStop}%
\bibitem [{\citenamefont {Morris}\ \emph {et~al.}(2020)\citenamefont {Morris},
  \citenamefont {Yadin}, \citenamefont {Fadel}, \citenamefont {Zibold},
  \citenamefont {Treutlein},\ and\ \citenamefont {Adesso}}]{2020Morris}%
  \BibitemOpen
  \bibfield  {author} {\bibinfo {author} {\bibfnamefont {B.}~\bibnamefont
  {Morris}}, \bibinfo {author} {\bibfnamefont {B.}~\bibnamefont {Yadin}},
  \bibinfo {author} {\bibfnamefont {M.}~\bibnamefont {Fadel}}, \bibinfo
  {author} {\bibfnamefont {T.}~\bibnamefont {Zibold}}, \bibinfo {author}
  {\bibfnamefont {P.}~\bibnamefont {Treutlein}},\ and\ \bibinfo {author}
  {\bibfnamefont {G.}~\bibnamefont {Adesso}},\ }\href
  {https://doi.org/10.1103/PhysRevX.10.041012} {\bibfield  {journal} {\bibinfo
  {journal} {Phys. Rev. X}\ }\textbf {\bibinfo {volume} {10}},\ \bibinfo
  {pages} {041012} (\bibinfo {year} {2020})}\BibitemShut {NoStop}%
\bibitem [{\citenamefont {Nosrati}\ \emph {et~al.}(2020)\citenamefont
  {Nosrati}, \citenamefont {Castellini}, \citenamefont {Compagno},\ and\
  \citenamefont {Lo~Franco}}]{2020Farzam}%
  \BibitemOpen
  \bibfield  {author} {\bibinfo {author} {\bibfnamefont {F.}~\bibnamefont
  {Nosrati}}, \bibinfo {author} {\bibfnamefont {A.}~\bibnamefont {Castellini}},
  \bibinfo {author} {\bibfnamefont {G.}~\bibnamefont {Compagno}},\ and\
  \bibinfo {author} {\bibfnamefont {R.}~\bibnamefont {Lo~Franco}},\ }\href
  {https://doi.org/10.1103/PhysRevA.102.062429} {\bibfield  {journal} {\bibinfo
   {journal} {Phys. Rev. A}\ }\textbf {\bibinfo {volume} {102}},\ \bibinfo
  {pages} {062429} (\bibinfo {year} {2020})}\BibitemShut {NoStop}%
\bibitem [{\citenamefont {Fr\'{e}rot}\ \emph {et~al.}(2023)\citenamefont
  {Fr\'{e}rot}, \citenamefont {Fadel},\ and\ \citenamefont
  {Lewenstein}}]{2023Frerot}%
  \BibitemOpen
  \bibfield  {author} {\bibinfo {author} {\bibfnamefont {I.}~\bibnamefont
  {Fr\'{e}rot}}, \bibinfo {author} {\bibfnamefont {M.}~\bibnamefont {Fadel}},\
  and\ \bibinfo {author} {\bibfnamefont {M.}~\bibnamefont {Lewenstein}},\
  }\href {https://doi.org/10.1088/1361-6633/acf8d7} {\bibfield  {journal}
  {\bibinfo  {journal} {Reports on Progress in Physics}\ }\textbf {\bibinfo
  {volume} {86}},\ \bibinfo {pages} {114001} (\bibinfo {year}
  {2023})}\BibitemShut {NoStop}%
\bibitem [{\citenamefont {Ptaszy\ifmmode~\acute{n}\else \'{n}\fi{}ski}\ and\
  \citenamefont {Esposito}(2023)}]{2023Ptaszynski}%
  \BibitemOpen
  \bibfield  {author} {\bibinfo {author} {\bibfnamefont {K.}~\bibnamefont
  {Ptaszy\ifmmode~\acute{n}\else \'{n}\fi{}ski}}\ and\ \bibinfo {author}
  {\bibfnamefont {M.}~\bibnamefont {Esposito}},\ }\href
  {https://doi.org/10.1103/PhysRevLett.130.150201} {\bibfield  {journal}
  {\bibinfo  {journal} {Phys. Rev. Lett.}\ }\textbf {\bibinfo {volume} {130}},\
  \bibinfo {pages} {150201} (\bibinfo {year} {2023})}\BibitemShut {NoStop}%
\bibitem [{\citenamefont {Serrano-Ens\'astiga}\ \emph
  {et~al.}(2024)\citenamefont {Serrano-Ens\'astiga}, \citenamefont {Denis},\
  and\ \citenamefont {Martin}}]{2024Serrano}%
  \BibitemOpen
  \bibfield  {author} {\bibinfo {author} {\bibfnamefont {E.}~\bibnamefont
  {Serrano-Ens\'astiga}}, \bibinfo {author} {\bibfnamefont {J.}~\bibnamefont
  {Denis}},\ and\ \bibinfo {author} {\bibfnamefont {J.}~\bibnamefont
  {Martin}},\ }\href {https://doi.org/10.1103/PhysRevA.109.022430} {\bibfield
  {journal} {\bibinfo  {journal} {Phys. Rev. A}\ }\textbf {\bibinfo {volume}
  {109}},\ \bibinfo {pages} {022430} (\bibinfo {year} {2024})}\BibitemShut
  {NoStop}%
\bibitem [{\citenamefont {Ajzenberg-Selove}(1988)}]{88Ajzen}%
  \BibitemOpen
  \bibfield  {author} {\bibinfo {author} {\bibfnamefont {F.}~\bibnamefont
  {Ajzenberg-Selove}},\ }\href
  {https://doi.org/http://dx.doi.org/10.1016/0375-9474(88)90124-8} {\bibfield
  {journal} {\bibinfo  {journal} {Nuclear Physics A}\ }\textbf {\bibinfo
  {volume} {490}},\ \bibinfo {pages} {1 } (\bibinfo {year} {1988})},\ \bibinfo
  {note} {note: several versions with the same title has been
  published.}\BibitemShut {Stop}%
\bibitem [{\citenamefont {Ajzenberg-Selove}(1991)}]{91Ajzen}%
  \BibitemOpen
  \bibfield  {author} {\bibinfo {author} {\bibfnamefont {F.}~\bibnamefont
  {Ajzenberg-Selove}},\ }\href
  {https://doi.org/http://dx.doi.org/10.1016/0375-9474(91)90446-D} {\bibfield
  {journal} {\bibinfo  {journal} {Nuclear Physics A}\ }\textbf {\bibinfo
  {volume} {523}},\ \bibinfo {pages} {1 } (\bibinfo {year} {1991})},\ \bibinfo
  {note} {with revised version at
  http://www.tunl.duke.edu/nucldata/.}\BibitemShut {Stop}%
\bibitem [{\citenamefont {Bochkarev}\ \emph {et~al.}(1989)\citenamefont
  {Bochkarev}, \citenamefont {Chulkov}, \citenamefont {Korsheninniicov},
  \citenamefont {in}, \citenamefont {Mukha},\ and\ \citenamefont
  {Yankov}}]{1989Boch}%
  \BibitemOpen
  \bibfield  {author} {\bibinfo {author} {\bibfnamefont {O.}~\bibnamefont
  {Bochkarev}}, \bibinfo {author} {\bibfnamefont {L.}~\bibnamefont {Chulkov}},
  \bibinfo {author} {\bibfnamefont {A.}~\bibnamefont {Korsheninniicov}},
  \bibinfo {author} {\bibfnamefont {E.~K.}\ \bibnamefont {in}}, \bibinfo
  {author} {\bibfnamefont {I.}~\bibnamefont {Mukha}},\ and\ \bibinfo {author}
  {\bibfnamefont {G.}~\bibnamefont {Yankov}},\ }\href
  {https://doi.org/http://dx.doi.org/10.1016/0375-9474(89)90371-0} {\bibfield
  {journal} {\bibinfo  {journal} {Nuclear Physics A}\ }\textbf {\bibinfo
  {volume} {505}},\ \bibinfo {pages} {215 } (\bibinfo {year}
  {1989})}\BibitemShut {NoStop}%
\bibitem [{\citenamefont {Grigorenko}\ \emph
  {et~al.}(2009{\natexlab{a}})\citenamefont {Grigorenko}, \citenamefont
  {Wiser}, \citenamefont {Mercurio}, \citenamefont {Charity}, \citenamefont
  {Shane}, \citenamefont {Sobotka}, \citenamefont {Elson}, \citenamefont
  {Wuosmaa}, \citenamefont {Banu}, \citenamefont {McCleskey}, \citenamefont
  {Trache}, \citenamefont {Tribble},\ and\ \citenamefont {Zhukov}}]{09Gri_80}%
  \BibitemOpen
  \bibfield  {author} {\bibinfo {author} {\bibfnamefont {L.~V.}\ \bibnamefont
  {Grigorenko}}, \bibinfo {author} {\bibfnamefont {T.~D.}\ \bibnamefont
  {Wiser}}, \bibinfo {author} {\bibfnamefont {K.}~\bibnamefont {Mercurio}},
  \bibinfo {author} {\bibfnamefont {R.~J.}\ \bibnamefont {Charity}}, \bibinfo
  {author} {\bibfnamefont {R.}~\bibnamefont {Shane}}, \bibinfo {author}
  {\bibfnamefont {L.~G.}\ \bibnamefont {Sobotka}}, \bibinfo {author}
  {\bibfnamefont {J.~M.}\ \bibnamefont {Elson}}, \bibinfo {author}
  {\bibfnamefont {A.~H.}\ \bibnamefont {Wuosmaa}}, \bibinfo {author}
  {\bibfnamefont {A.}~\bibnamefont {Banu}}, \bibinfo {author} {\bibfnamefont
  {M.}~\bibnamefont {McCleskey}}, \bibinfo {author} {\bibfnamefont
  {L.}~\bibnamefont {Trache}}, \bibinfo {author} {\bibfnamefont {R.~E.}\
  \bibnamefont {Tribble}},\ and\ \bibinfo {author} {\bibfnamefont {M.~V.}\
  \bibnamefont {Zhukov}},\ }\href {https://doi.org/10.1103/PhysRevC.80.034602}
  {\bibfield  {journal} {\bibinfo  {journal} {Phys. Rev. C}\ }\textbf {\bibinfo
  {volume} {80}},\ \bibinfo {pages} {034602} (\bibinfo {year}
  {2009}{\natexlab{a}})}\BibitemShut {NoStop}%
\bibitem [{\citenamefont {Grigorenko}\ \emph
  {et~al.}(2009{\natexlab{b}})\citenamefont {Grigorenko}, \citenamefont
  {Wiser}, \citenamefont {Miernik}, \citenamefont {Charity}, \citenamefont
  {Pftzner}, \citenamefont {Banu}, \citenamefont {Bingham}, \citenamefont
  {Cwiok}, \citenamefont {Darby}, \citenamefont {Dominik}, \citenamefont
  {Elson}, \citenamefont {Ginter}, \citenamefont {Grzywacz}, \citenamefont
  {Janas}, \citenamefont {Karny}, \citenamefont {Korgul}, \citenamefont
  {Liddick}, \citenamefont {Mercurio}, \citenamefont {Rajabali}, \citenamefont
  {Rykaczewski}, \citenamefont {Shane}, \citenamefont {Sobotka}, \citenamefont
  {Stolz}, \citenamefont {Trache}, \citenamefont {Tribble}, \citenamefont
  {Wuosmaa},\ and\ \citenamefont {Zhukov}}]{09Gri_677}%
  \BibitemOpen
  \bibfield  {author} {\bibinfo {author} {\bibfnamefont {L.}~\bibnamefont
  {Grigorenko}}, \bibinfo {author} {\bibfnamefont {T.}~\bibnamefont {Wiser}},
  \bibinfo {author} {\bibfnamefont {K.}~\bibnamefont {Miernik}}, \bibinfo
  {author} {\bibfnamefont {R.}~\bibnamefont {Charity}}, \bibinfo {author}
  {\bibfnamefont {M.}~\bibnamefont {Pftzner}}, \bibinfo {author} {\bibfnamefont
  {A.}~\bibnamefont {Banu}}, \bibinfo {author} {\bibfnamefont {C.}~\bibnamefont
  {Bingham}}, \bibinfo {author} {\bibfnamefont {M.}~\bibnamefont {Cwiok}},
  \bibinfo {author} {\bibfnamefont {I.}~\bibnamefont {Darby}}, \bibinfo
  {author} {\bibfnamefont {W.}~\bibnamefont {Dominik}}, \bibinfo {author}
  {\bibfnamefont {J.}~\bibnamefont {Elson}}, \bibinfo {author} {\bibfnamefont
  {T.}~\bibnamefont {Ginter}}, \bibinfo {author} {\bibfnamefont
  {R.}~\bibnamefont {Grzywacz}}, \bibinfo {author} {\bibfnamefont
  {Z.}~\bibnamefont {Janas}}, \bibinfo {author} {\bibfnamefont
  {M.}~\bibnamefont {Karny}}, \bibinfo {author} {\bibfnamefont
  {A.}~\bibnamefont {Korgul}}, \bibinfo {author} {\bibfnamefont
  {S.}~\bibnamefont {Liddick}}, \bibinfo {author} {\bibfnamefont
  {K.}~\bibnamefont {Mercurio}}, \bibinfo {author} {\bibfnamefont
  {M.}~\bibnamefont {Rajabali}}, \bibinfo {author} {\bibfnamefont
  {K.}~\bibnamefont {Rykaczewski}}, \bibinfo {author} {\bibfnamefont
  {R.}~\bibnamefont {Shane}}, \bibinfo {author} {\bibfnamefont
  {L.}~\bibnamefont {Sobotka}}, \bibinfo {author} {\bibfnamefont
  {A.}~\bibnamefont {Stolz}}, \bibinfo {author} {\bibfnamefont
  {L.}~\bibnamefont {Trache}}, \bibinfo {author} {\bibfnamefont
  {R.}~\bibnamefont {Tribble}}, \bibinfo {author} {\bibfnamefont
  {A.}~\bibnamefont {Wuosmaa}},\ and\ \bibinfo {author} {\bibfnamefont
  {M.}~\bibnamefont {Zhukov}},\ }\href
  {https://doi.org/http://dx.doi.org/10.1016/j.physletb.2009.04.085} {\bibfield
   {journal} {\bibinfo  {journal} {Physics Letters B}\ }\textbf {\bibinfo
  {volume} {677}},\ \bibinfo {pages} {30 } (\bibinfo {year}
  {2009}{\natexlab{b}})}\BibitemShut {NoStop}%
\bibitem [{\citenamefont {Egorova}\ \emph {et~al.}(2012)\citenamefont
  {Egorova}, \citenamefont {Charity}, \citenamefont {Grigorenko}, \citenamefont
  {Chajecki}, \citenamefont {Coupland}, \citenamefont {Elson}, \citenamefont
  {Ghosh}, \citenamefont {Howard}, \citenamefont {Iwasaki}, \citenamefont
  {Kilburn}, \citenamefont {Lee}, \citenamefont {Lynch}, \citenamefont
  {Manfredi}, \citenamefont {Marley}, \citenamefont {Sanetullaev},
  \citenamefont {Shane}, \citenamefont {Shetty}, \citenamefont {Sobotka},
  \citenamefont {Tsang}, \citenamefont {Winkelbauer}, \citenamefont {Wuosmaa},
  \citenamefont {Youngs},\ and\ \citenamefont {Zhukov}}]{12Ego}%
  \BibitemOpen
  \bibfield  {author} {\bibinfo {author} {\bibfnamefont {I.~A.}\ \bibnamefont
  {Egorova}}, \bibinfo {author} {\bibfnamefont {R.~J.}\ \bibnamefont
  {Charity}}, \bibinfo {author} {\bibfnamefont {L.~V.}\ \bibnamefont
  {Grigorenko}}, \bibinfo {author} {\bibfnamefont {Z.}~\bibnamefont
  {Chajecki}}, \bibinfo {author} {\bibfnamefont {D.}~\bibnamefont {Coupland}},
  \bibinfo {author} {\bibfnamefont {J.~M.}\ \bibnamefont {Elson}}, \bibinfo
  {author} {\bibfnamefont {T.~K.}\ \bibnamefont {Ghosh}}, \bibinfo {author}
  {\bibfnamefont {M.~E.}\ \bibnamefont {Howard}}, \bibinfo {author}
  {\bibfnamefont {H.}~\bibnamefont {Iwasaki}}, \bibinfo {author} {\bibfnamefont
  {M.}~\bibnamefont {Kilburn}}, \bibinfo {author} {\bibfnamefont
  {J.}~\bibnamefont {Lee}}, \bibinfo {author} {\bibfnamefont {W.~G.}\
  \bibnamefont {Lynch}}, \bibinfo {author} {\bibfnamefont {J.}~\bibnamefont
  {Manfredi}}, \bibinfo {author} {\bibfnamefont {S.~T.}\ \bibnamefont
  {Marley}}, \bibinfo {author} {\bibfnamefont {A.}~\bibnamefont {Sanetullaev}},
  \bibinfo {author} {\bibfnamefont {R.}~\bibnamefont {Shane}}, \bibinfo
  {author} {\bibfnamefont {D.~V.}\ \bibnamefont {Shetty}}, \bibinfo {author}
  {\bibfnamefont {L.~G.}\ \bibnamefont {Sobotka}}, \bibinfo {author}
  {\bibfnamefont {M.~B.}\ \bibnamefont {Tsang}}, \bibinfo {author}
  {\bibfnamefont {J.}~\bibnamefont {Winkelbauer}}, \bibinfo {author}
  {\bibfnamefont {A.~H.}\ \bibnamefont {Wuosmaa}}, \bibinfo {author}
  {\bibfnamefont {M.}~\bibnamefont {Youngs}},\ and\ \bibinfo {author}
  {\bibfnamefont {M.~V.}\ \bibnamefont {Zhukov}},\ }\href
  {https://doi.org/10.1103/PhysRevLett.109.202502} {\bibfield  {journal}
  {\bibinfo  {journal} {Phys. Rev. Lett.}\ }\textbf {\bibinfo {volume} {109}},\
  \bibinfo {pages} {202502} (\bibinfo {year} {2012})}\BibitemShut {NoStop}%
\bibitem [{\citenamefont {Oishi}\ \emph {et~al.}(2014)\citenamefont {Oishi},
  \citenamefont {Hagino},\ and\ \citenamefont {Sagawa}}]{2014Oishi}%
  \BibitemOpen
  \bibfield  {author} {\bibinfo {author} {\bibfnamefont {T.}~\bibnamefont
  {Oishi}}, \bibinfo {author} {\bibfnamefont {K.}~\bibnamefont {Hagino}},\ and\
  \bibinfo {author} {\bibfnamefont {H.}~\bibnamefont {Sagawa}},\ }\href
  {https://doi.org/10.1103/PhysRevC.90.034303} {\bibfield  {journal} {\bibinfo
  {journal} {Phys. Rev. C}\ }\textbf {\bibinfo {volume} {90}},\ \bibinfo
  {pages} {034303} (\bibinfo {year} {2014})}\BibitemShut {NoStop}%
\bibitem [{\citenamefont {Oishi}\ \emph {et~al.}(2017)\citenamefont {Oishi},
  \citenamefont {Kortelainen},\ and\ \citenamefont {Pastore}}]{2017Oishi}%
  \BibitemOpen
  \bibfield  {author} {\bibinfo {author} {\bibfnamefont {T.}~\bibnamefont
  {Oishi}}, \bibinfo {author} {\bibfnamefont {M.}~\bibnamefont {Kortelainen}},\
  and\ \bibinfo {author} {\bibfnamefont {A.}~\bibnamefont {Pastore}},\ }\href
  {https://doi.org/10.1103/PhysRevC.96.044327} {\bibfield  {journal} {\bibinfo
  {journal} {Phys. Rev. C}\ }\textbf {\bibinfo {volume} {96}},\ \bibinfo
  {pages} {044327} (\bibinfo {year} {2017})}\BibitemShut {NoStop}%
\bibitem [{\citenamefont {Wang}\ and\ \citenamefont
  {Nazarewicz}(2021)}]{2021Wang_Naza}%
  \BibitemOpen
  \bibfield  {author} {\bibinfo {author} {\bibfnamefont {S.~M.}\ \bibnamefont
  {Wang}}\ and\ \bibinfo {author} {\bibfnamefont {W.}~\bibnamefont
  {Nazarewicz}},\ }\href {https://doi.org/10.1103/PhysRevLett.126.142501}
  {\bibfield  {journal} {\bibinfo  {journal} {Phys. Rev. Lett.}\ }\textbf
  {\bibinfo {volume} {126}},\ \bibinfo {pages} {142501} (\bibinfo {year}
  {2021})}\BibitemShut {NoStop}%
\bibitem [{\citenamefont {Cirel'son}(1980)}]{1980Tsirelson}%
  \BibitemOpen
  \bibfield  {author} {\bibinfo {author} {\bibfnamefont {B.~S.}\ \bibnamefont
  {Cirel'son}},\ }\href {https://doi.org/10.1007/BF00417500} {\bibfield
  {journal} {\bibinfo  {journal} {Letters in Mathematical Physics}\ }\textbf
  {\bibinfo {volume} {4}},\ \bibinfo {pages} {93} (\bibinfo {year}
  {1980})}\BibitemShut {NoStop}%
\bibitem [{\citenamefont {Nielsen}\ and\ \citenamefont
  {Chuang}(2000)}]{2000Nielsen}%
  \BibitemOpen
  \bibfield  {author} {\bibinfo {author} {\bibfnamefont {M.}~\bibnamefont
  {Nielsen}}\ and\ \bibinfo {author} {\bibfnamefont {I.}~\bibnamefont
  {Chuang}},\ }\href@noop {} {\emph {\bibinfo {title} {Quantum Computation and
  Quantum Information}}}\ (\bibinfo  {publisher} {Cambridge University Press},\
  \bibinfo {address} {Cambridge, UK},\ \bibinfo {year} {2000})\BibitemShut
  {NoStop}%
\bibitem [{\citenamefont {Edmonds}(1960)}]{60Edm}%
  \BibitemOpen
  \bibfield  {author} {\bibinfo {author} {\bibfnamefont {A.~R.}\ \bibnamefont
  {Edmonds}},\ }\href@noop {} {\emph {\bibinfo {title} {Angular Momentum in
  Quantum Mechanics}}},\ Princeton Landmarks in Physics\ (\bibinfo  {publisher}
  {Princeton University Press},\ \bibinfo {address} {Princeton, USA},\ \bibinfo
  {year} {1960})\BibitemShut {NoStop}%
\bibitem [{\citenamefont {Suzuki}\ and\ \citenamefont
  {Ikeda}(1988)}]{88Suzuki_COSM}%
  \BibitemOpen
  \bibfield  {author} {\bibinfo {author} {\bibfnamefont {Y.}~\bibnamefont
  {Suzuki}}\ and\ \bibinfo {author} {\bibfnamefont {K.}~\bibnamefont {Ikeda}},\
  }\href {https://doi.org/10.1103/PhysRevC.38.410} {\bibfield  {journal}
  {\bibinfo  {journal} {Phys. Rev. C}\ }\textbf {\bibinfo {volume} {38}},\
  \bibinfo {pages} {410} (\bibinfo {year} {1988})}\BibitemShut {NoStop}%
\bibitem [{\citenamefont {Bertsch}\ and\ \citenamefont
  {Esbensen}(1991)}]{1991BE}%
  \BibitemOpen
  \bibfield  {author} {\bibinfo {author} {\bibfnamefont {G.}~\bibnamefont
  {Bertsch}}\ and\ \bibinfo {author} {\bibfnamefont {H.}~\bibnamefont
  {Esbensen}},\ }\href
  {https://doi.org/http://dx.doi.org/10.1016/0003-4916(91)90033-5} {\bibfield
  {journal} {\bibinfo  {journal} {Annals of Physics}\ }\textbf {\bibinfo
  {volume} {209}},\ \bibinfo {pages} {327 } (\bibinfo {year}
  {1991})}\BibitemShut {NoStop}%
\bibitem [{\citenamefont {Esbensen}\ \emph {et~al.}(1997)\citenamefont
  {Esbensen}, \citenamefont {Bertsch},\ and\ \citenamefont
  {Hencken}}]{1997EBH}%
  \BibitemOpen
  \bibfield  {author} {\bibinfo {author} {\bibfnamefont {H.}~\bibnamefont
  {Esbensen}}, \bibinfo {author} {\bibfnamefont {G.~F.}\ \bibnamefont
  {Bertsch}},\ and\ \bibinfo {author} {\bibfnamefont {K.}~\bibnamefont
  {Hencken}},\ }\href {https://doi.org/10.1103/PhysRevC.56.3054} {\bibfield
  {journal} {\bibinfo  {journal} {Phys. Rev. C}\ }\textbf {\bibinfo {volume}
  {56}},\ \bibinfo {pages} {3054} (\bibinfo {year} {1997})}\BibitemShut
  {NoStop}%
\bibitem [{\citenamefont {Hagino}\ and\ \citenamefont {Sagawa}(2005)}]{2005HS}%
  \BibitemOpen
  \bibfield  {author} {\bibinfo {author} {\bibfnamefont {K.}~\bibnamefont
  {Hagino}}\ and\ \bibinfo {author} {\bibfnamefont {H.}~\bibnamefont
  {Sagawa}},\ }\href {https://doi.org/10.1103/PhysRevC.72.044321} {\bibfield
  {journal} {\bibinfo  {journal} {Phys. Rev. C}\ }\textbf {\bibinfo {volume}
  {72}},\ \bibinfo {pages} {044321} (\bibinfo {year} {2005})}\BibitemShut
  {NoStop}%
\bibitem [{\citenamefont {Thompson}\ \emph {et~al.}(1977)\citenamefont
  {Thompson}, \citenamefont {Lemere},\ and\ \citenamefont {Tang}}]{77Thom}%
  \BibitemOpen
  \bibfield  {author} {\bibinfo {author} {\bibfnamefont {D.}~\bibnamefont
  {Thompson}}, \bibinfo {author} {\bibfnamefont {M.}~\bibnamefont {Lemere}},\
  and\ \bibinfo {author} {\bibfnamefont {Y.}~\bibnamefont {Tang}},\ }\href
  {https://doi.org/http://dx.doi.org/10.1016/0375-9474(77)90007-0} {\bibfield
  {journal} {\bibinfo  {journal} {Nuclear Physics A}\ }\textbf {\bibinfo
  {volume} {286}},\ \bibinfo {pages} {53 } (\bibinfo {year}
  {1977})}\BibitemShut {NoStop}%
\bibitem [{\citenamefont {Horodecki}\ \emph {et~al.}(2009)\citenamefont
  {Horodecki}, \citenamefont {Horodecki}, \citenamefont {Horodecki},\ and\
  \citenamefont {Horodecki}}]{2009Horod}%
  \BibitemOpen
  \bibfield  {author} {\bibinfo {author} {\bibfnamefont {R.}~\bibnamefont
  {Horodecki}}, \bibinfo {author} {\bibfnamefont {P.}~\bibnamefont
  {Horodecki}}, \bibinfo {author} {\bibfnamefont {M.}~\bibnamefont
  {Horodecki}},\ and\ \bibinfo {author} {\bibfnamefont {K.}~\bibnamefont
  {Horodecki}},\ }\href {https://doi.org/10.1103/RevModPhys.81.865} {\bibfield
  {journal} {\bibinfo  {journal} {Rev. Mod. Phys.}\ }\textbf {\bibinfo {volume}
  {81}},\ \bibinfo {pages} {865} (\bibinfo {year} {2009})}\BibitemShut
  {NoStop}%
\end{thebibliography}


\end{document}